\documentclass{article}
\usepackage[utf8]{inputenc}
\usepackage{geometry}
\usepackage[table]{xcolor}
\usepackage{hyperref}
\usepackage{multirow}
\usepackage{graphicx}
\usepackage{booktabs,subcaption, longtable, array, float}
\usepackage{anyfontsize}
\usepackage{dirtytalk}
\usepackage{amsmath,amssymb}
\usepackage{lscape}
\usepackage{adjustbox}
\usepackage{authblk}
\usepackage{threeparttable}
\usepackage[normalem]{ulem}
\usepackage[backend=biber, style=numeric, citestyle=numeric, sorting=none]{biblatex}
\graphicspath{{Figures/}}

\renewcommand{\title}[1]{{\noindent\centering\Large\bfseries#1\medskip\\[1em]}}
\renewcommand{\author}[2]{{\noindent\large #1 \medskip\\[0.5em] \noindent \small #2 \medskip\\}}

\begin{document}

\title{Evacuation patterns and socioeconomic stratification in the context of wildfires in Chile}

\author{T. Naushirvanov,\textsuperscript{1}
E. Elejalde,\textsuperscript{2}
K. Kalimeri,\textsuperscript{3}
E. Omodei,\textsuperscript{1}
M. Karsai,\textsuperscript{1,4}
L. Ferres\textsuperscript{3,5}
}
{
1. Department of Network and Data Science, Central European University, Vienna, Austria\\
2. L3S Research Center, Hannover, Germany\\
3. ISI Foundation, Turin, Italy\\
4. National Laboratory for Health Security, HUN-REN R\'enyi Institute of Mathematics, Hungary \\
5. IDS UDD, Santiago, Chile
}

\setlength{\extrarowheight}{1em}

\begin{abstract}
Climate change is altering the frequency and intensity of wildfires, leading to increased evacuation events that disrupt human mobility and socioeconomic structures. These disruptions affect access to resources, employment, and housing, amplifying existing vulnerabilities within communities. Understanding the interplay between climate change, wildfires, evacuation patterns, and socioeconomic factors is crucial for developing effective mitigation and adaptation strategies. To contribute to this challenge, we use high-definition mobile phone records to analyse evacuation patterns during the wildfires in Valparaíso, Chile, that took place between February 2-3, 2024. This data allows us to track the movements of individuals in the disaster area, providing insight into how people respond to large-scale evacuations in the context of severe wildfires. We apply a causal inference approach that combines regression discontinuity and difference-in-differences methodologies to observe evacuation behaviours during wildfires, with a focus on socioeconomic stratification. This approach allows us to isolate the impact of the wildfires on different socioeconomic groups by comparing the evacuation patterns of affected populations before and after the event, while accounting for underlying trends and discontinuities at the threshold of the disaster. We find that many people spent nights away from home, with those in the lowest socioeconomic segment stayed away the longest. In general, people reduced their travel distance during the evacuation, and the lowest socioeconomic group moved the least. Initially, movements became more random, as people sought refuge in a rush, but eventually gravitated towards areas with similar socioeconomic status. Our results show that socioeconomic differences play a role in evacuation dynamics, providing useful insights for response planning.
\end{abstract}


\section*{Introduction}


The increasing frequency and severity of natural disasters are part of a broader trend related to climate change that has become evident in recent years~\cite{xu_global_2023, walsh_extreme_2020, jolly_climate-induced_2015}.
As global temperatures rise, the incidence of wildfires is expected to increase, forcing governments and international organisations to reassess and improve their response strategies. 
Beyond changing weather conditions, human activities, such as the construction of settlements and infrastructure near flammable vegetation, have also intensified the frequency of wildfires~\cite{bowman_vegetation_2020}.
The impact of wildfires extends beyond immediate physical and economic damage. It also leads to increased exposure to air pollution, which affects the health of people, especially those in the low-income strata \cite{xu_global_2023}, and contributes to the significant release of greenhouse gases, with the potential to further accelerate global warming~\cite{wang_severe_2024}. 
Therefore, response plans should not only enhance the adaptive capacities of communities, especially in areas prone to climate-related disasters, but also address broader challenges related to health and forest management.

Such an event unfolded on the night of February 2-3, 2024, when severe wildfires ravaged the well-known and touristic area of Valparaíso, Chile, marking the country's worst natural disaster since the earthquakes in 2010, and the most devastating wildfire in the past 30 years.
This resulted in 137 deaths, left 1,600 people homeless, and directly affected more than 16,000 people~\cite{cursino_chile_2024}.
In response, the Chilean Government declared a State of Emergency and Catastrophe, and the Health Ministry issued a Sanitarian Alert, leading to curfews and the deployment of additional firefighting and rescue teams. 
Further, the Ministry of Health responded by hiring medical students to alleviate pressure on the healthcare system~\cite{france24_pictures_2024}.
Despite all these efforts, the emergency response faced numerous challenges; the wildfire caused extensive damage to drinking water supply systems and severely affected the health situation. A UNICEF humanitarian report highlighted the importance of delivering the necessary supplies and services to the most affected population, the majority of whom were children and adolescents ~\cite{unicef_emergency_team_unicef_2024}. Many residents were forced to evacuate from their homes, seeking shelter elsewhere. These challenges highlighted vulnerable points in actual response policies and underscored the need for improved emergency strategies.

Telecommunication and GPS data are invaluable tools to follow-up on such rapid behavioural changes, enabling real-time, high-resolution observations for a large number of individuals.
For instance, in the 2019 Sonoma wildfires evacuation process, researchers used GPS data to systematically analyse and identify different groups of evacuees ~\cite{zhao_estimating_2022}. 
Similar data have been used to develop a knowledge database to store evacuation plans for typical population distributions, significantly accelerating the process of finding near-optimal evacuation plans for urban emergency management~\cite{yin_improving_2020}. Estimation of real-time population movements during large disasters has been carried out using various kinds of mobile phone data and data assimilation techniques, combined with simulation of population movement and observation data, to estimate prediction accuracy and find ways to improve it~\cite{sekimoto_real-time_2016}.  GPS data was used in a study conducted on evacuation behaviours during four earthquakes in Japan to show that an individual’s evacuation probability depends on the intensity of the seismic they experience~\cite{yabe_cross-comparative_2019}. At the same time, the distance travelled during evacuation does not appear to depend on the intensity of the seismic event itself.

In addition to telecommunication and GPS data, social media data have become valuable tools for disaster monitoring. 
Facebook Disaster Maps (FBDM) have been used to study evacuation mobility patterns during the two mega-fires in California, USA, in 2018~\cite{jia_patterns_2020}. In that study, the FBDM was found to be representative of the California population, and analysis 
indicated three stages in evacuation mobility dynamics: a drastic decrease after the evacuation order, a significant increase near open shelters or nearby towns, and a gradual return to baseline after the lifting of the evacuation order. 

Diverse adaptation capacities to emergency scenarios following a natural disaster due to living, employment, and financial conditions can affect groups with divergent socioeconomic characteristics differently. Through understanding of such inequalities, effective response strategies can be developed to compensate for the various impact effects during emergencies.
Social vulnerabilities have been found to significantly impact evacuation decisions during wildfires, with notable differences between geographic areas~\cite{sun_social_2024}. In the same study, by analysing individual characteristics, unemployment emerged as a critical factor that negatively influences both the timing of evacuation and the distance traveled to their evacuation destinations. However, the impact of social vulnerabilities (such as being impoverished or non-white) on evacuation rates varied significantly across different census block groups, and their effects on departure delays and destination distances were found to be more uniform. 
Furthermore, evacuees with higher income were found to be more likely to evacuate from affected areas and reach safer locations with less damage to housing and infrastructure~\cite{yabe_effects_2020}. These differences were common among evacuees within and outside mandatory evacuation zones. Meanwhile the trends of population recovery after a displacement were found to be quite homogeneous among different socioeconomic groups~\cite{yabe_understanding_2020}. Similarly, community resilience, defined as a function of the magnitude of impact and recovery time, was assessed from GPS data during Hurricane Harvey, uncovering pronounced socioeconomic and racial disparities in evacuation and recovery patterns~\cite{hong_measuring_2021}. Racial and wealth disparities have been found to be important in evacuation patterns, with disadvantaged minorities being less likely to evacuate than wealthier Caucasian residents~\cite{deng_high-resolution_2021}.

In this paper, we investigate the impact of natural disasters on post-crisis human behaviour, in the highly segregated context of Chile. 
By analysing the Valparaíso wildfire as a case study, our aim was to understand its implications on the behaviour of different socioeconomic groups of the population to observe adoption capacity differences. In addition, we compare our observations to similar data collected by FBDM to validate aggregated disaster maps when confronted with more precise individual-level data from mobile phone records. 


\section*{Methods}\label{sec:methods}

For our investigations, we use anonymous data provided by a major mobile phone operator in South America (Telefónica Movistar\footnote{Hereafter mainly referred to as the mobile data provider.}), with a market share of 27\% in 2023~\cite{telefonica_consolidated_2024}. The observed mobile phone population moderately correlates with the official population at the census zone level ($\rho = .36$, see Supplementary Figure~\ref{fig:plot11_1}), a fine-grained intermediate area between the block and the census district\footnote{User manual for the database of the 2017 population and housing census, Department of Demography and Censuses, National Institute of Statistics of Chile, September 2018 (file in Spanish). Available at: \href{https://redatam-ine.ine.cl/manuales/Manual-Usuario.pdf}{this link.}}.

We analyse data from two time periods, 11 to 17 November 2023 and 19 January to 19 February 2024. The periods 11 to 17 November and 19 January to 1 February are considered baselines of normal, business-as-usual human mobility. Instead, the period from January 1 to February 19 covers the days after the wildfire outbreak. 


To determine socioeconomic status in the context of the lack of accurate self-reported socioeconomic attributes, we assign an approximate socioeconomic profile to the individual according to their inferred home location.
However, in the literature, there is little consensus on the optimal criteria to implement when creating decision rules for home detection methods~\cite{vanhoof_assessing_2018}.
\citeauthor{pappalardo_evaluation_2021} having thoroughly evaluated 37 home location algorithms, concluded that the most efficient approach based on CDR records is the Simple Matching Coefficient (SMC) \cite{bojic_choosing_2015} estimated between 7 pm and 7 am for approximately two weeks \cite{pappalardo_evaluation_2021}.  

We infer individuals' home locations by estimating for each of them their most visited tower during the night (from 12.00 am to 05.00 am), a much more conservative measure than usual \cite{pappalardo_evaluation_2021}. We assume that the tower is an individual's home location if (1) it is their most visited tower for at least six nights between 19 January and 1 February 2024, and (2) it is their most visited tower for at least five nights during a ``business-as-usual" week, between 11 and 17 November 2023. These conditions help to ensure that the individuals included in our sample are part of a local population that can be seriously affected by the wildfire. Using this approach, we were able to identify 282,118 unique phone IDs present in both periods. Of these, 126,129 unique phone IDs were successfully assigned a home location and for 115,600 (91.65\%) the home location assignments were the same for both periods.

During the analysis of changes in human behaviour, we differentiate among three groups of individuals. 
\begin{itemize}
    \item \textbf{Potentially affected} are those who spent one night from January 31 to February 2 near towers within a 5-kilometre radius around the areas warned (towers that received a warning about the wildfire and a need to evacuate).
    \item \textbf{Likely evacuated} is a subgroup of potentially affected with a home location within the affected area and who were away from their home tower at least once during the nights from January 31 to February 4.  
    \item \textbf{Not affected} are individuals observed outside the 5 km warned areas on the corresponding nights and used as a control group for comparison.
\end{itemize}

Following our definition of potentially affected, likely evacuated and not affected, we have 156,896 potentially affected unique phone IDs and 200,851 not affected ones. Of them, we could identify stable home locations for 47,487 and 54,637 unique phone IDs, accordingly. Potentially affected people include 28,676 unique IDs of likely evacuated people (60.38\% of potentially affected with inferred home locations).

Having inferred the home location of the individuals in the dataset, we assign them an estimated socioeconomic status corresponding to the average socioeconomic status of the zone according to the census. We use the percentage of people with higher education as a proxy for socioeconomic status and divide districts into three socioeconomic groups: Low, Medium, and High.
We consider a socioeconomic division based on quantiles; each socioeconomic quantile (bin) has the same number of individuals living in each census zone. Table~\ref{tab:ses} shows the distribution of identified mobile phone users in each sociodemographic group according to their home location.

\begin{table}[H]
\centering
\caption{SES Distribution of Unique Mobile Devices Among Affected Groups}
\label{tab:ses}
\begin{tabular}{|c|c|c|c|c|}
\hline
\textbf{Group} & \textbf{Low} & \textbf{Medium} & \textbf{High} \\ \hline
Potentially Affected & 12,334 & 18,929 & 16,224 \\ \hline
Likely Evacuated & 7,059 & 11,226 & 10,391 \\ \hline
Not Affected & 19,646 & 17,549 & 17,441 \\ \hline
\end{tabular}
\end{table}

\subsection*{Causal Modelling}
\textbf{Regression Discontinuity in Time}

We applied a Regression Discontinuity in Time (RDiT) design to assess the causal effects of people's travel patterns before and after the wildfires~\cite{hausman_regression_2018}. In this context, treatment refers to compensating for the impact of wildfires that affected the local population and potentially forced people to relocate. Our RDiT design is based on the assumption that, in the absence of treatment, evacuated people would continue their daily routine movement and there would be no noticeable discontinuity. Since time cannot be assigned randomly, another traditional RDiT assumption of local randomisation cannot work in this scenario. To account for heteroskedasticity and autocorrelation, the Newey-West variance estimator is used.
Our model is defined as:

\[
\text{Y}_{t} = \alpha + \beta \cdot \Delta_t + \gamma \cdot \text{threshold}_{t} + \delta (\Delta_t \times \text{threshold}_{t}) + \theta \cdot \text{Controls}_{t} + \epsilon_{t}, 
\]
where \(\text{Y}_{t}\) is the dependent variable of choice, typically representing either the fraction of people who leave their home towers or the average distance travelled (in km) at time \(t\). \(\Delta_t\) represents time (in days) relative to the threshold, which is defined as the wildfire event that provoked evacuation, occurring during the night from February 2 to February 3. The term \(\text{threshold}_{t}\) is an indicator variable that equals 1 if time \(t\) is after the threshold. The coefficient \(\beta\) captures the effect of time on the dependent variable of choice, while \(\gamma\) represents the immediate effect of the wildfire event (threshold) on the dependent variable. The interaction term \(\delta\) indicates how the effect of time changes after the threshold. The term \(\theta \cdot \text{controls}_{t}\) represents control variables, specifically accounting for weekday effects, which capture differences in the dependent variable based on the day of the week. Finally, \(\epsilon_{t}\) is the error term.

Further modifications include socioeconomic classes and their interaction with the threshold value. This interaction term indicates changes in a dependent variable of choice corresponding to a particular socioeconomic class after the beginning of the wildfire.

\newpage

\begin{figure}[t]
    \centering
    \includegraphics[width=\linewidth]{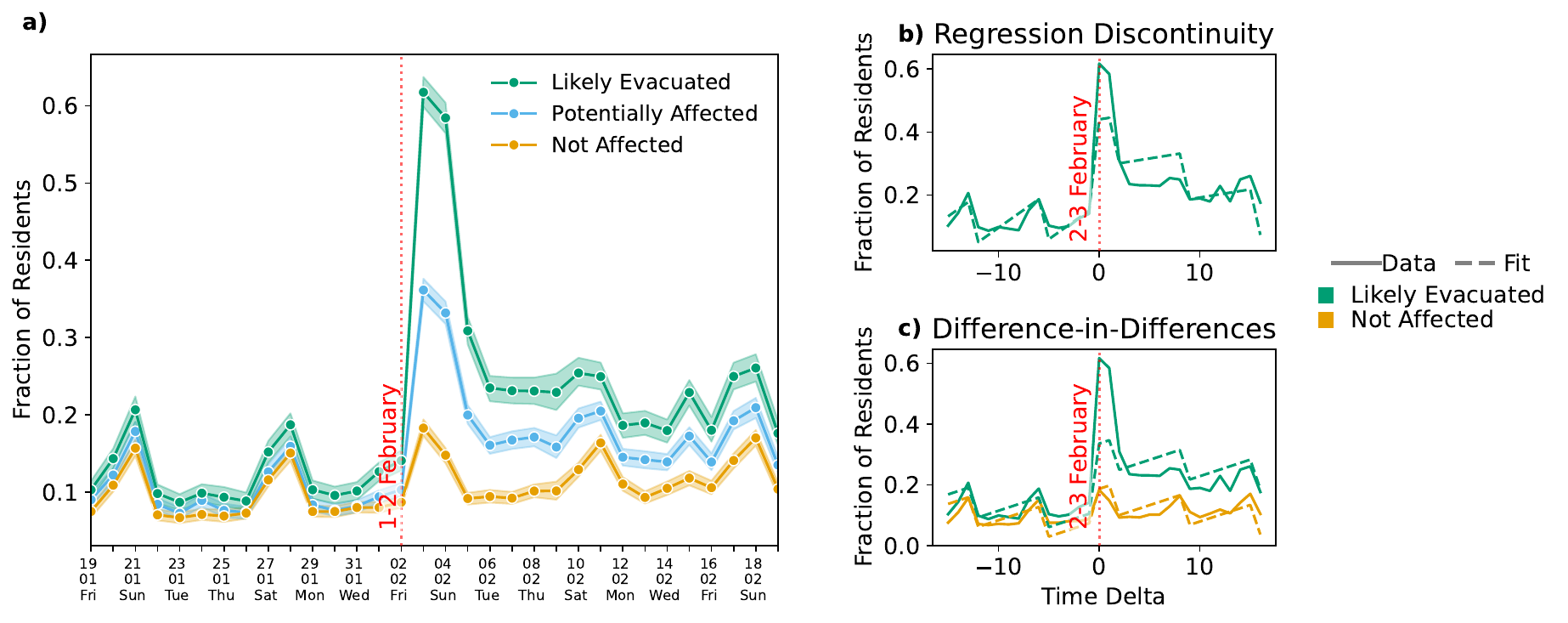}
    \caption{\textbf{Fraction of individuals whose night location was different from their home location.} \textbf{a)} Fractions of individuals by three group types (Not Affected, Potentially Affected, and Likely Evacuated), with a 95\% confidence interval. 95\% CI obtained through bootstrapping the fraction of moved people by iteratively resampling the original dataset (1,000 times) with replacement, each time generating a sample representing around 10\% of the population. \textbf{b)} Fractions of likely evacuated people as given in data and inferred with the regression discontinuity in time model. \textbf{c)} Fractions of likely evacuated and not affected people as given in data and inferred with the difference in differences model.}
    \label{fig:plot1}
\end{figure}

\noindent \textbf{Difference-in-Differences}

The second approach we employ is the Difference-in-Differences (DiD) method, which is used to estimate causal effects by comparing changes in outcomes over time between a treatment group and a control group, before and after an intervention~\cite{wing_designing_2018}. In our case, the treatment group are those people who were evacuated due to wildfires, while the control group is the non-affected population, according to our definitions above. The DiD design relies on the parallel trend assumption that, in the absence of intervention (wildfires), both the treatment and control groups would behave the same. We also assume that there are no spillover effects: the treatment effect is confined to the treated group and does not affect the control group, which is reasonable taking into account our definitions of evacuated, affected, and non-affected populations. DiD analysis helps control for factors that change over time but are not related to treatment, assuming that these factors affect both the treatment and control groups in the same way. Our model is defined as:

\[
\text{Y}_{t} = \alpha + \beta \cdot \text{threshold}_{t} + \gamma \cdot \text{treatment}_{t} + \delta (\text{threshold}_{t} \times \text{treatment}_{t}) + \theta \cdot \text{Controls}_{t} + \epsilon_{t},
\]
where \(\text{Y}_{t}\) represents the dependent variable of choice, usually the fraction of people away from their home towers or the average distance travelled (in km) at time \(t\). The term \(\text{threshold}_{t}\) is an indicator variable equal to 1 if time \(t\) is after the threshold event. \(\text{treatment}_{t}\) is an indicator variable equal to 1 for evacuated people (treatment group) and 0 for non-affected people (control group). The coefficient \(\beta\) captures the effect of time on the dependent variable of choice, while \(\gamma\) represents the difference in the dependent variable between the treatment and control groups before the wildfire. The term \(\delta\) is the Difference-in-Differences estimator, showing the differential effect of the wildfire on the treatment group relative to the control group. The \(\theta \cdot \text{Controls}_{t}\) term represents control variables, including weekday effects (which capture the differences in the dependent variable based on the day of the week) and \(\Delta_t\) (which represents time in days relative to the threshold). Finally, \(\epsilon_{t}\) is the error term.

As in the case with RDiT, additional modifications also include socioeconomic classes and their interaction with the treatment and threshold variables. This "triple" difference-in-differences indicates changes in a dependent variable of choice corresponding to a particular socioeconomic class of the likely affected people after the beginning of the wildfire compared to those not affected.

Using RDiT and DiD together, we show how wildfire (intervention) affected population groups differently based on socioeconomic status, providing insight into the effectiveness of emergency responses and the disparities in impacts on different populations. Using the coefficient \(\gamma\) near the threshold term in the RDiT design, we can compare changes in displacement patterns exclusively for likely evacuated people. In DiD, the main coefficient of interest is \(\delta\) near the difference-in-differences interaction term, as it additionally compares changes in displacement patterns relative to the control group of not affected people.

\vspace{-0.1cm}
\section*{Results}\label{sec:results}

\vspace{-0.1cm}
\subsection*{Measuring evacuation rates and travelled distances}

\begin{figure}[t]
    \centering
    \includegraphics[width=\linewidth]{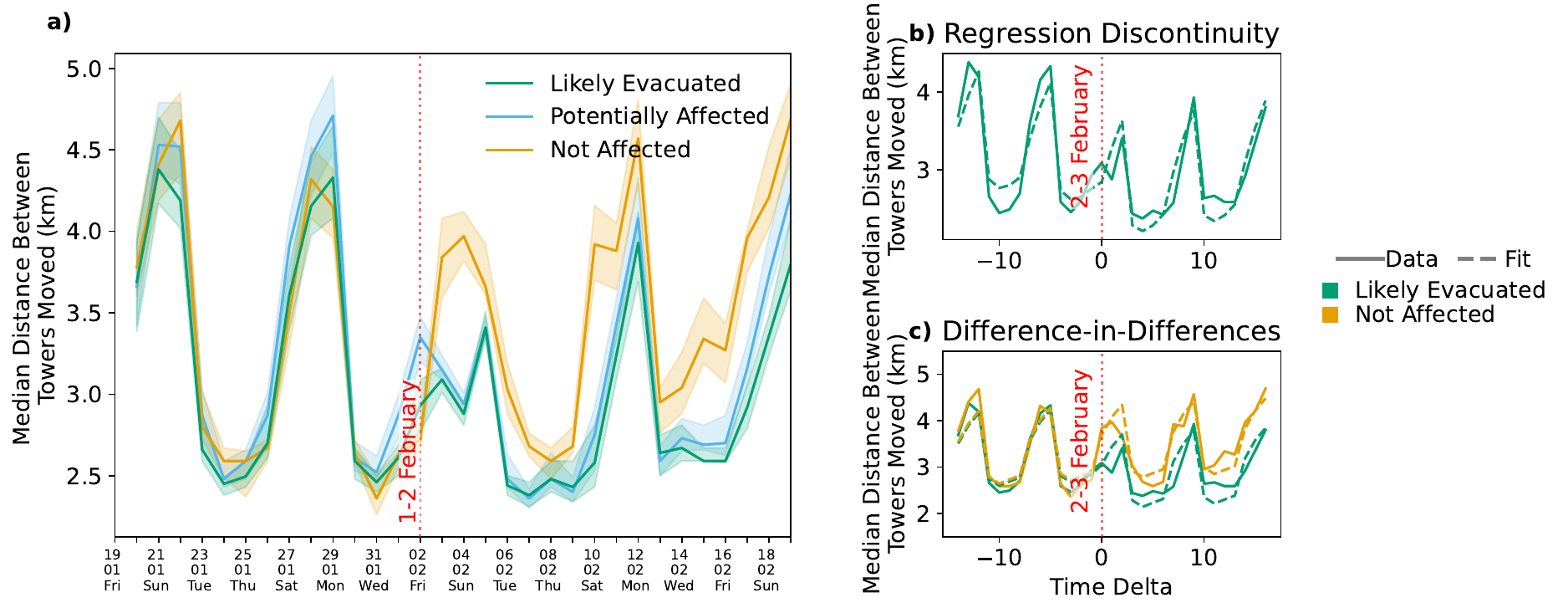}
    \caption{\textbf{Median travel distance between the towers (only for individuals who moved to another tower).} \textbf{a)} Median travel distances by three group types (Not Affected, Potentially Affected, and Likely Evacuated), with a 95\% confidence interval. \textbf{b)} Median travel distances of likely evacuated people as given in data and inferred with the regression discontinuity in time model. \textbf{c)} Median travel distances of likely evacuated and not affected people as given in data and inferred with the difference in differences model.}
    \label{fig:plot2}
\end{figure}

To assess behavioural differences between affected and unaffected groups, we analyse aggregated patterns of human behaviour after the onset of wildfires in Chile. Initially, we look at the fraction of people who evacuated from their home\footnote{Hereinafter, ``home" refers to the inferred home tower.} on the night of the fire. Figure \ref{fig:plot1} shows a clear difference between the three types of populations observed. Before the wildfire, the percentage of people away from home rarely exceeded 20\% among all the groups analysed. However, after the night when the wildfires struck, this percentage increased for likely evacuated people to more than 60\%, while the trend for non-affected people remained more or less the same. This difference remains clear between the three groups for the weeks following the natural disaster.

Both regression discontinuity (Supplementary Table \ref{tab:regression_results_fraction_rdd}) and difference-in-differences (Supplementary Table \ref{tab:regression_results_fraction_dd}) models showed a statistically significant increase in the fraction of likely evacuated people who had to spend nights away from their home, especially compared to unaffected people. For example, after the onset of the wildfires, the fraction of evacuated individuals spending a night away increased by an average of 0.267, compared to the same set of people before the wildfire (all else being equal), or an average of 0.118, when compared to the control group of non-affected (all else being equal). This result highlights the substantial impact of the wildfires on the displacement patterns of the affected population, which, in turn, supports the validity and precision of our methodology in identifying target populations.

In addition, we explore the distances between the most visited towers every night for each individual before and after the beginning of the wildfire. Figure \ref{fig:plot2} shows the median travel distances for those individuals who moved to a different tower at night. Despite seeing before that the fraction of evacuated people who spent a night away noticeably increased, the median travelling distance of the evacuated people in kilometres rather decreased compared to the previous weeks and compared to the non-affected people. This suggests that although more people having moved due to the wildfire, they did not go that far from their home, rather choosing relatively close areas (e.g. shelters, relatives, and friends) for relocation.

When applying regression discontinuity and difference-in-differences, we can identify a statistically significant decrease of 0.42 km for evacuated people, compared to their average travel distances before the wildfires (Supplementary Table \ref{tab:regression_results_median_rdd}), or a reduction of 0.60 km compared to non-affected people (Supplementary Table \ref{tab:regression_results_median_dd}). This finding suggests that more people tended to seek refuge in relatively nearby locations rather than travelling long distances.

\vspace{-0.1cm}
\subsection*{Socioeconomic differences in displacement patterns}

\begin{figure}[t]
    \centering
    \begin{adjustbox}{width=1\textwidth,center}
        \includegraphics{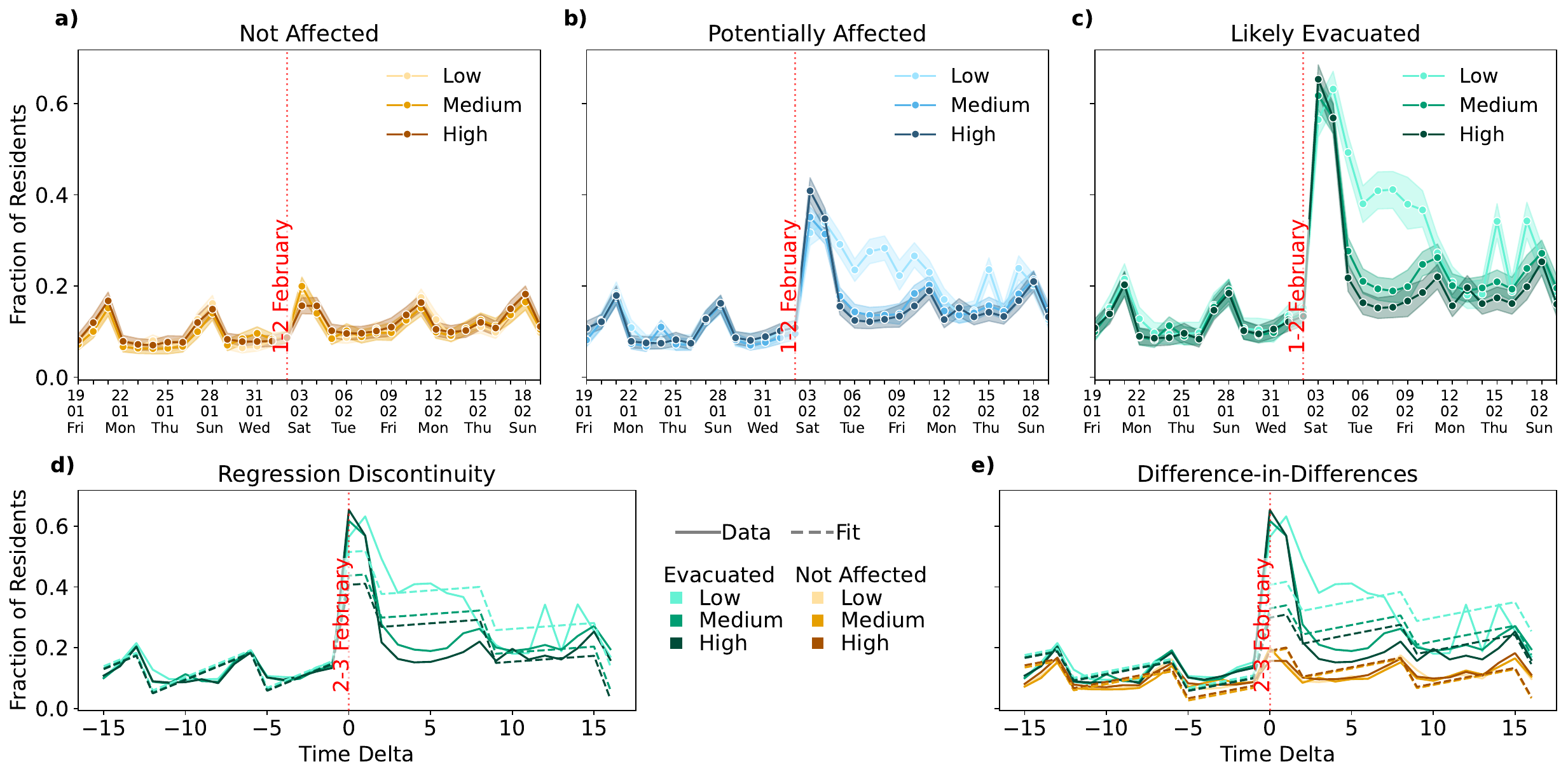}
    \end{adjustbox}
    \caption{
          \textbf{Fraction of individuals whose night location was different from their home location by socioeconomic status.} \textbf{a)-c)} Fractions of individuals by three group types (Not Affected, Potentially Affected, and Likely Evacuated) and by socioeconomic status (Low, Medium, High), with a 95\% confidence interval. 95\% CI obtained through bootstrapping the fraction of moved people by iteratively resampling the original dataset (1,000 times) with replacement, each time generating a sample representing around 10\% of the population. \textbf{d)} Fractions of likely evacuated people (by socioeconomic status) as given in data and inferred with the regression discontinuity in time model. \textbf{e)} Fractions of likely evacuated and not affected people (by socioeconomic status) as given in data and inferred with the difference-in-differences model.
    }
    \label{fig:plot3}
\end{figure}

After identifying general trends and differences between likely evacuated and not affected populations, we can examine further variations in the behaviours of different socioeconomic groups. Due to a lack of accurate self-reported socioeconomic attributes, we assign an approximate socioeconomic profile to the individual according to their inferred home location. As the inferred home location corresponds to a census zone, we divided available census zones into three socioeconomic groups (Low, Medium, and High) using the percentage of people with higher education as a socioeconomic proxy. Each socioeconomic group contains a similar number of individuals.

\begin{figure}[t]
    \centering
    \begin{adjustbox}{width=1\textwidth,center}
        \includegraphics{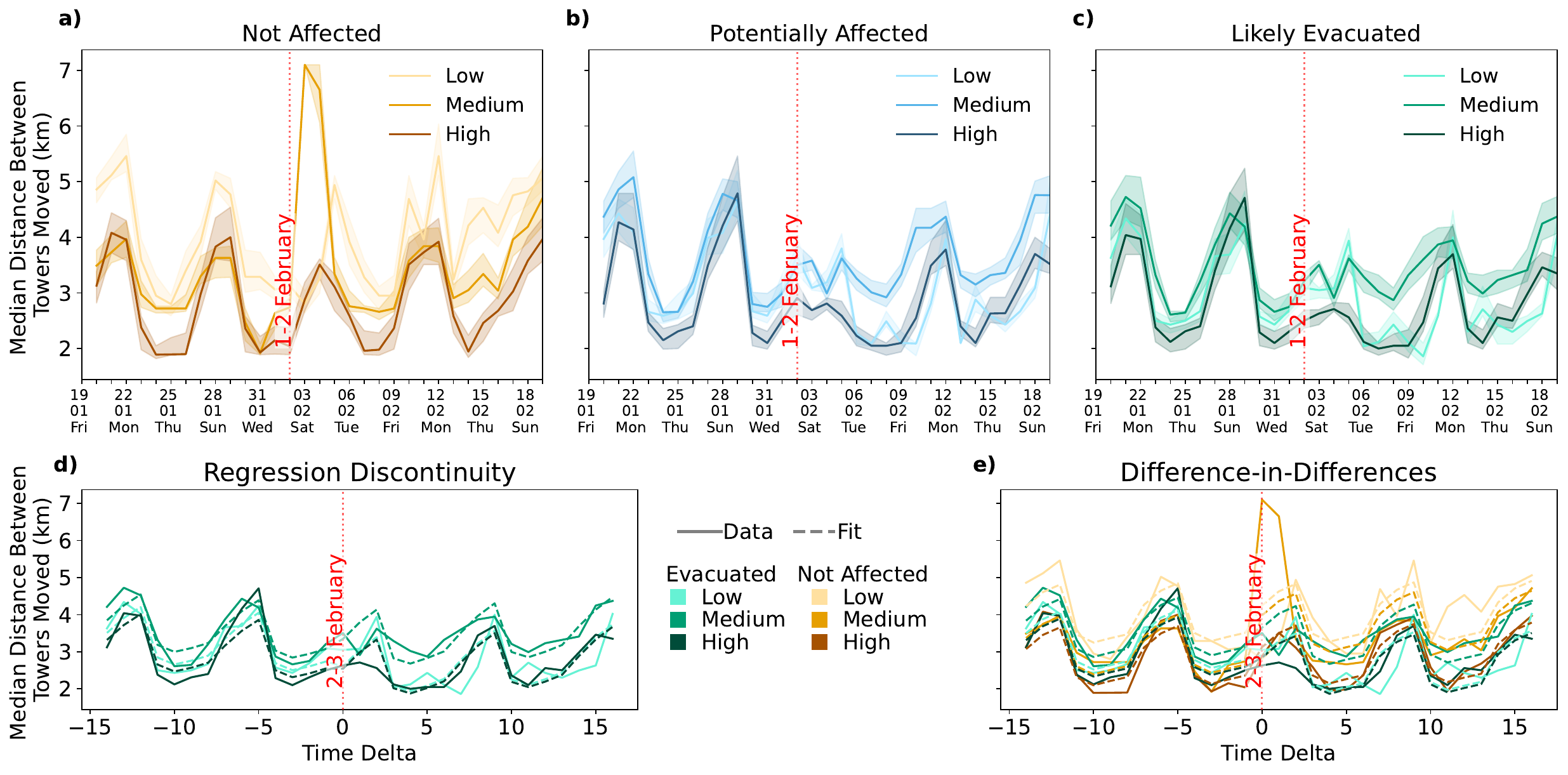}
    \end{adjustbox}
    \caption{
          \textbf{Median travel distance between the towers by socioeconomic status (only for individuals who moved to another tower).} \textbf{a)-c)} Median travel distances of individuals by three group types (Not Affected, Potentially Affected, and Likely Evacuated) and by socioeconomic status (Low, Medium, High), with a 95\% confidence interval. \textbf{d)} Median travel distances of likely evacuated people (by socioeconomic status) as given in data and inferred with the regression discontinuity in time model. \textbf{e)} Median travel distances of likely evacuated and not affected people (by socioeconomic status) as given in data and inferred with the difference-in-differences model.}
    \label{fig:plot4}
\end{figure}

Figure \ref{fig:plot3} illustrates the differences in trends among the three socioeconomic groups. Although there are no noticeable differences between these groups for the unaffected population before or after the wildfires, the variations for the potentially affected and likely evacuated populations are significant. Firstly, the people in the lowest socioeconomic group stayed away from their homes for a longer period. The same is true for the medium socioeconomic class compared to the richest, although this difference is considerably smaller. Secondly, the highest proportion of the lowest socioeconomic class was reached the day after similar peaks occurred for people of medium and high socioeconomic classes. This may imply that certain people from lower socioeconomic areas needed more time to adapt to the crisis.

Regression models confirmed significant statistical differences in behavioural responses between different socioeconomic groups after the onset of wildfires. For example, compared to previous time periods, the fraction of evacuated individuals from the medium and high socioeconomic classes who had to spend a night away from their home towers was lower by 0.070 and 0.098, respectively, compared to the low socioeconomic class (Supplementary Table \ref{tab:regression_results_fraction_rdd}). This difference between socioeconomic groups was not statistically significant before the natural disaster occurred. Similarly, compared to the control group of non-affected people, these fractions were lower by 0.072 and 0.097 for the medium and high socioeconomic classes, respectively (Supplementary Table \ref{tab:regression_results_fraction_dd}).

When we look at the median kilometres travelled by individuals from different socioeconomic classes (Figure \ref{fig:plot4}), it appears that there is a greater variation between socioeconomic classes even within the non-affected sample. Regressions confirm this statistically significant difference, showing that, compared to people of lower socioeconomic status, people of the rich and middle classes move on average a shorter distance, by 1.153 and 0.857 km, respectively (Supplementary Table \ref{tab:regression_results_median_dd})~\footnote{Given that the wildfire occurred near the city of Valparaíso, and the entire affected area is quite urbanised, a distance difference of 1.1 km is substantial in this context.}. However, once we focus on the impact of the wildfire on mobility, we still see a significant decrease in travel distances for the likely evacuated people. Nevertheless, in this case, there are no statistically significant variations between the socioeconomic groups in the likely-evacuated group compared to the control group. After the beginning of the natural disaster, people from all socioeconomic groups experienced a similar decrease in their median travel distances, moving from one tower to another. At the same time, if we focus only on the movement of those likely evacuated before the wildfire, we can see a greater statistically significant decrease for the low and high socioeconomic groups (without significant differences between the two), while the movement of the medium socioeconomic class was less affected (Supplementary Table \ref{tab:regression_results_median_dd}). It appears that the movement of people from the lowest and richest socioeconomic groups was more constrained in space than that of the middle-class areas.

\subsection*{Investigating segregation in displacement}

\begin{figure}[t]
    \centering
    \begin{adjustbox}{width=1\textwidth,center}
    \includegraphics{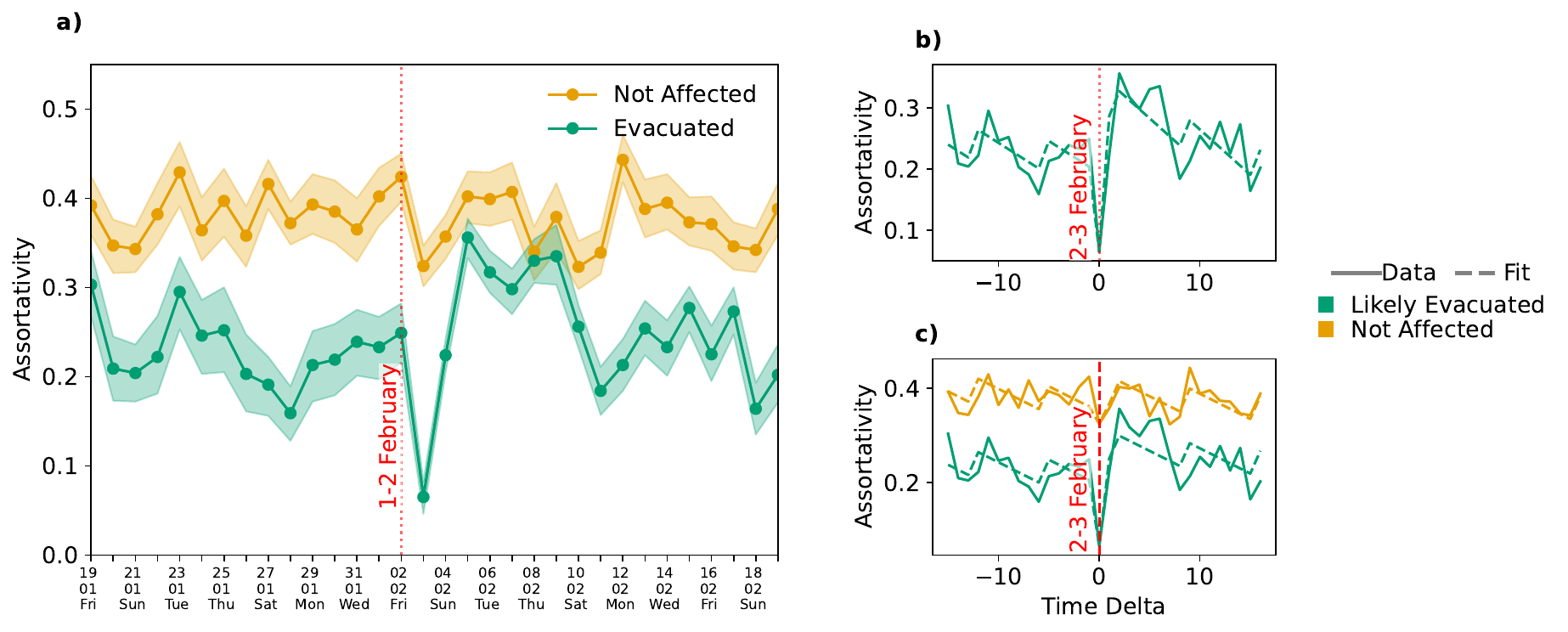}
    \end{adjustbox}
    \caption{
        \textbf{Assortativity of people who changed night towers.} \textbf{a)} Assortativity over time for two groups of people (Not Affected and Likely Evacuated), with a 95\% confidence interval. The 95\% CI was obtained through bootstrapping the assortativity values of moved individuals by iteratively resampling the original dataset (1,000 times) with replacement. For each resample, an assortativity measure was computed using a heatmap representing movement patterns between socioeconomic categories, capturing the variability in assortativity over time. \textbf{b)} Assortativity of likely evacuated people as given in data and inferred with the regression discontinuity in time model. \textbf{c)} Assortativity of likely evacuated and not affected people as given in data and inferred with the difference-in-differences model.
    }
    \label{fig:plot5}
\end{figure}

To understand the effect of socioeconomic segregation on displacement patterns, we use the assortativity coefficient. The assortativity coefficient represents the correlation coefficient of stratification matrices, which is widely used in the study of human mobility patterns~\cite{bokanyi_universal_2021}. The stratification matrix illustrates the aggregated movement of individuals between areas of different or similar socioeconomic classes. Given the normalised matrix $\tilde{X}$, where the trips between $i$ and $j$ are normalised over the total trips that occur in the system, we calculate the \textit{assortativity} $\rho$ with the Pearson correlation coefficient of the matrix entries, across all income groups.

\[\rho = \frac{\sum_{i,j}ij \tilde{X}_{ij} - \sum_{i,j}i\tilde{X}_{ij}\sum_{i,j} j \tilde{X}_{ij}}
{
\sqrt{\sum_{i,j} i^2 \tilde{X}_{ij} - (\sum_{i,j} i \tilde{X}_{ij})^2}
\sqrt{\sum_{i, j} j^2\tilde{X}_{ij} - (\Sigma_{i,j} j \tilde{X}_{ij})^2}
}\]

Researchers have already shown that, in the context of urban mobility, people tend to visit places of the same socioeconomic class more often~\cite{bokanyi_universal_2021, napoli_socioeconomic_2023}. We use the assortativity coefficient to determine whether displacement patterns during wildfires exhibit similar segregation. A completely assortative matrix will have an assortativity value of $\rho = 1$, indicating that displacement is highly segregated, with individuals from similar socioeconomic backgrounds moving to the areas of the same socioeconomic status. In contrast, a completely disassortative matrix will have $\rho = -1$, suggesting that displacement is characterised by individuals from different socioeconomic groups moving to areas with completely different statuses. Figure \ref{fig:plot5} helps compare the differences in the assortativity values between evacuated and non-affected people.

After the beginning of the wildfire, there was a noticeable drop in assortativity for evacuated people, which is mainly due to the fact that people had to leave their home locations in haste (see SI, Supplementary Figure \ref{fig:plot5_all}). If we focus only on the assortativity of those who moved (Figure \ref{fig:plot5}), these values also showed a slight drop the first day after, but then reached higher values compared to previous weeks. For the non-affected people, these changes were not differentiable from the previous two weeks.

Regression modelling confirms the statistical significance of these changes. Compared to the control group, the assortativity of the likely evacuated people decreased by 0.143 on the first night of forced mobility changes (2-3 February), and their overall assortativity after the beginning of wildfires increased by 0.04 (all else being equal) (Supplementary Table \ref{tab:regression_results_assort_moved_dd}). Compared with its time trend before the crisis, assortativity decreased by 0.236 on the first night of forced mobility changes but generally increased by 0.108 during the crisis (Supplementary Table \ref{tab:regression_results_assort_moved_rdd}).

These results suggest that people likely forced to evacuate due to the wildfire initially relocated to areas with varying socioeconomic statuses, indicating that socioeconomic factors were not a significant determinant in their immediate choice of destination (see \cite{elejalde_social_2024} for a similar phenomenon in long-term relocation). However, after a few days, their movement assortativity increased. This pattern may indicate that evacuees eventually moved to stay with friends or family members, who are more likely to share a similar socioeconomic status.

\subsection*{Comparing mobile and social media data}

\begin{figure}[t]
    \centering
    \begin{adjustbox}{width=1\textwidth,center}
        \includegraphics{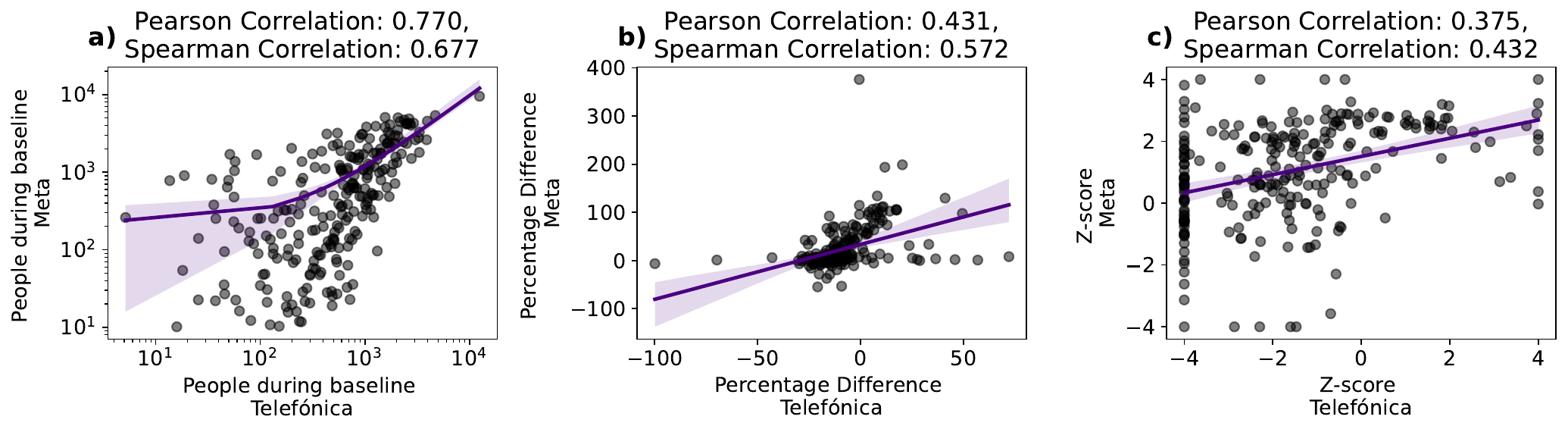}
    \end{adjustbox}
    \caption{
        \textbf{Correlations between three indicators in Telefónica and Meta datasets across geolocated cells for Thursday, February 8, at 13:00 (GMT-4).} \textbf{a)} Comparing recorded number of users between two datasets in pre-crisis period. \textbf{b)} Comparing percentage changes of users activity before and after the crisis between two datasets. \textbf{c)} Comparing changes in z-scores of users activity before and after the crisis between two datasets.}
    \label{fig:plot7}
\end{figure}

As a final step of this analysis, we compare Telefónica data with Facebook Population in Crisis Data~\cite{meta_facebook_nodate}. Meta collects crisis data for different types of disasters. These datasets show the number of Facebook users with geolocation enabled on a grid of approximately $2.4 \times 2.4$ km, in a time window of 8 hours, correcting for a baseline of Facebook usage before and after emergencies. A particularity of these datasets is that data collection begins after a disaster occurs, and hence the behaviour of the people prior to or during the disaster is not observed, unlike our data. The motivation behind our comparison is to cross-validate the use of Facebook and mobile phone data sets to analyse human displacement patterns during natural disasters and similar crises. This analysis is especially crucial because Meta's crisis data are readily accessible to practitioners and policymakers, providing them with valuable insights on population mobility and displacement.

\begin{figure}[t]
    \centering
    \begin{adjustbox}{width=1\textwidth,center}
        \includegraphics{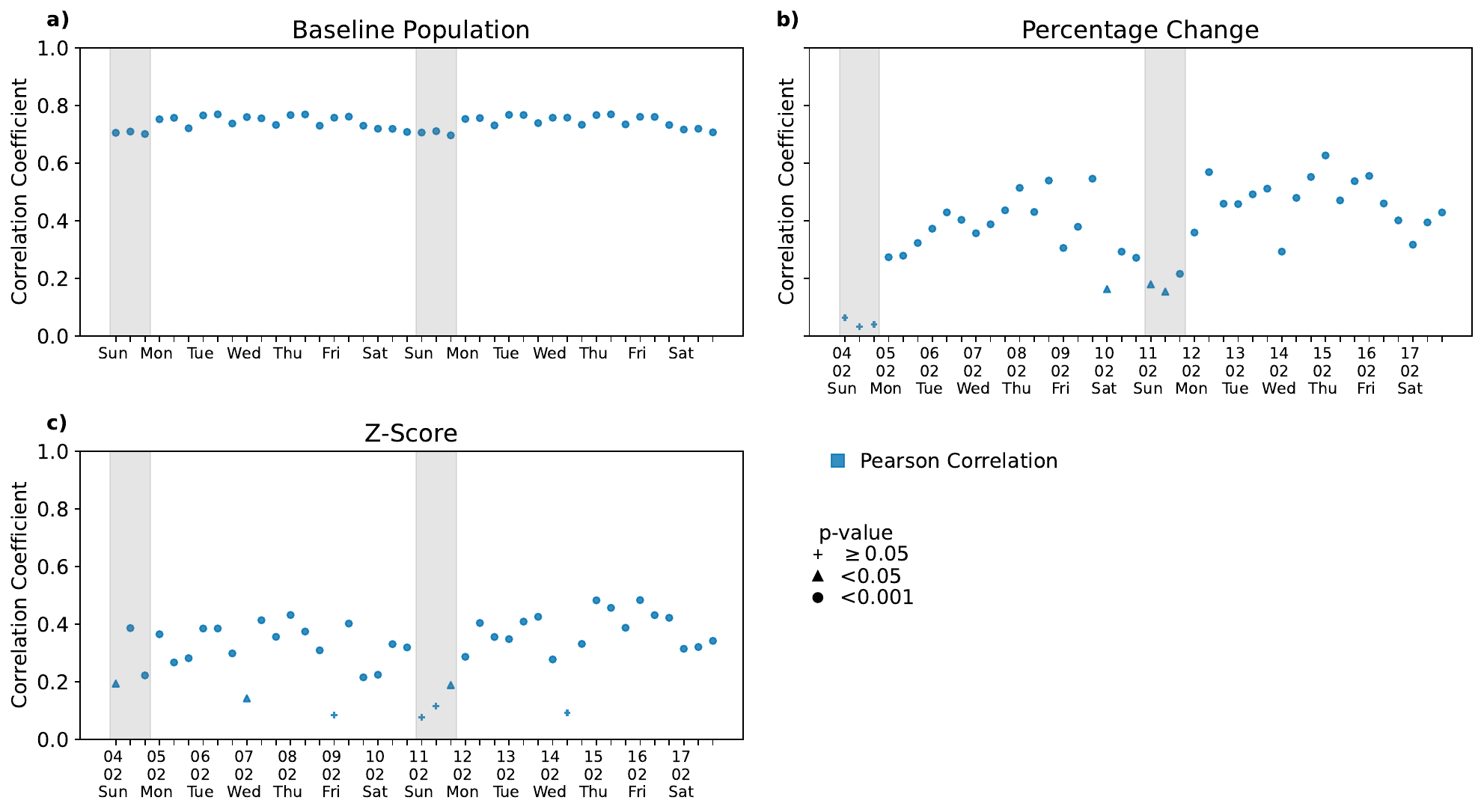}
    \end{adjustbox}
    \caption{
            \textbf{Pearson correlations between three indicators in Telefónica and Meta datasets across geolocated cells over time.} \textbf{a)} Comparing recorded number of users between two datasets in pre-crisis period over time. Sundays coloured grey.  \textbf{b)} Comparing percentage changes of users activity before and after the crisis between two datasets over time. Sundays coloured grey.  \textbf{c)} Comparing changes in z-scores of users activity before and after the crisis between two datasets over time. Sundays coloured grey. }
    \label{fig:plot8}
\end{figure}

To create a dataset comparable to that of Facebook's Population in Crisis, we calculated the number of unique mobile devices connected to each tower at 15-minute intervals. These data were also temporally aggregated into 8-hour intervals and spatially by the size of cells in the Facebook grid. A tower belongs to a particular Facebook cell if it is located inside this cell. If a tower does not belong to any cell, it is assigned to the nearest grid, provided the distance is no more than 10 km. Since Facebook data are available from February 4 to February 19, we use the same time frame for our dataset derived from mobile phone data. For the baseline values that indicate the number of people connected to each tower before the crisis, we use the average values from January 19 to February 1. These values were aggregated by weekday and 8-hour time intervals.

After this data transformation, we compare the Facebook and Telefónica datasets. First, we compare the correlations of different indicators in the two datasets: the average number of people during the pre-crisis period, the percentage changes in the number of people in the post-crisis period compared to the baseline, and the z-scores of these changes. Figure \ref{fig:plot7} shows the correlations for a specific time and date (Thursday, February 8, 13:00), and Figure \ref{fig:plot8} presents the Pearson correlation throughout the observation period. As can be seen, the highest correlation between both datasets is achieved when comparing population baselines, which are calculated as the number of geolocated Facebook users or the number of active phone IDs in the areas under investigation.

Additionally, both percentage change and z-score reflect changes in the number of people present in these areas compared to the pre-crisis period. We observe relatively strong correlations for percentage changes, although there are some periods, particularly on Sundays, when correlation values drop. The correlations of the z-score tend to be lower. In general, the correlation of both measurements tends to be variable; sometimes it is around 0.6, while at other times it drops below 0.1.

\begin{figure}[H]
    \centering
    \begin{adjustbox}{width=0.6\textwidth,center}
    \includegraphics{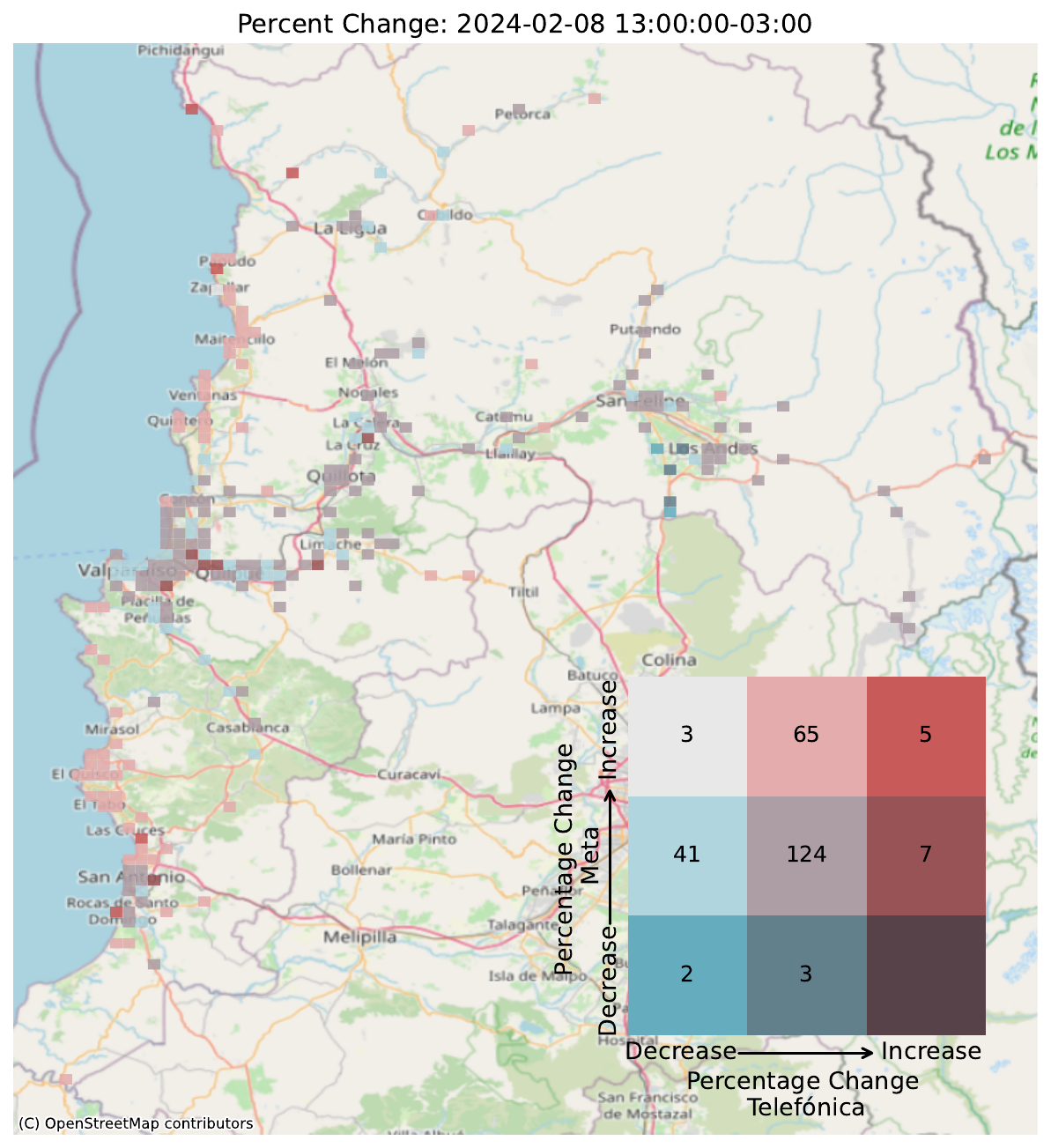}
    \end{adjustbox}
    \caption{
        \textbf{Confusion Matrix of recorded Percentage Changes for Thursday, February 8, at 13:00 (GMT-4).} Accuracy equals 0.52.
        }
    \label{fig:plot9_1}
\end{figure}

After analysing general correlations, we categorised each measurement (percentage change and z-score) into three groups and compared these categories between the Meta and mobile data datasets. These categories represent whether human activity around each tower increased, remained stable, or decreased during the crisis period. The three categories are: Increase (percentage change $\geq 20$ or z-score $\geq 2$), Stable (percentage change from -20 to 20 or z-score from -2 to 2, excluding border values), and Decrease (percentage change $\leq -20$ or z-score $\leq -2$). Figure \ref{fig:plot9_1} shows an example of a confusion matrix for February 8, 13:00. To compare classes between data from the two data providers, the following accuracy formula is used:

\begin{equation}
\text{Accuracy} = \frac{1}{N} \sum_{i=1}^{N} \mathbb{I}(y_i = x_i),
\end{equation}

where \( N \) represents the total number of geotiles, \( y_i \) is the Meta label for the \( i \)-th geotile, and \( x_i \) is the Telefónica label for the \( i \)-th geotile. The term \( \mathbb{I} \) is the indicator function, which takes the value 1 if the condition inside is true (i.e., \( y_i = x_i \)) and 0 otherwise.

\begin{figure}[H]
    \centering
    \begin{adjustbox}{width=0.7\textwidth,center}
    \includegraphics{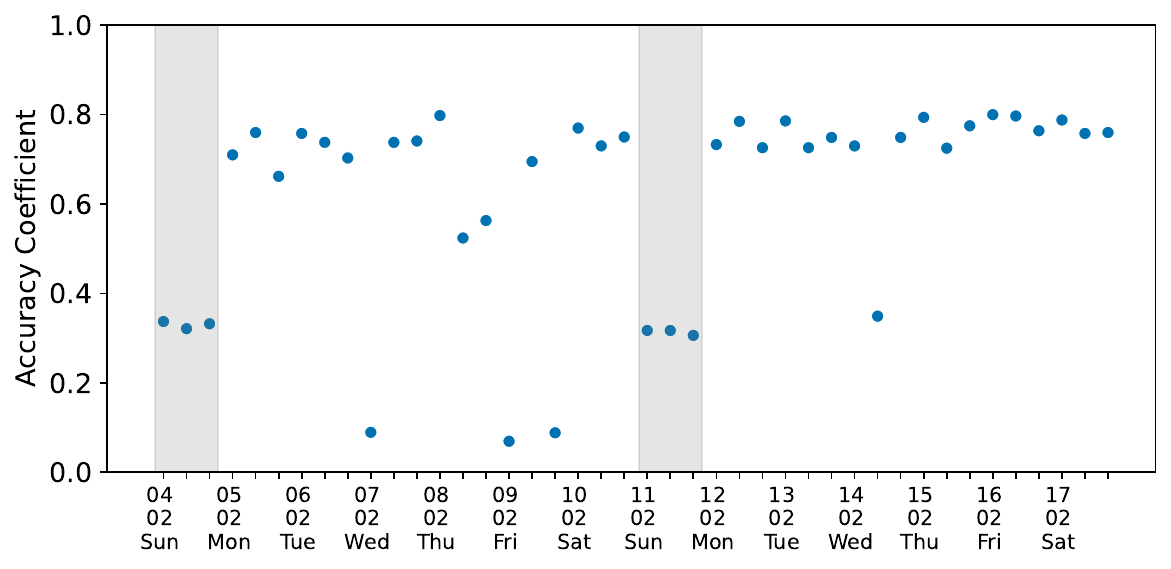}
    \end{adjustbox}
    \caption{
        \textbf{Accuracy scores for the Percentage Change in different time periods.} Sundays coloured grey.
        }
    \label{fig:plot9_2}
\end{figure}

The accuracy score varies from 0 (when there is no overlap between two datasets) to 1 (complete overlap). The confusion matrix score in Figure \ref{fig:plot9_1} is equal to 0.52, indicating that more than half of the geolocated tiles showed similar mobility patterns in both datasets. Figure \ref{fig:plot9_2} shows the accuracy score of Categorised Percentage Change over time (for changes in z-score, see SI, Supplementary Figures \ref{fig:plot13_1} and \ref{fig:plot13_2}). The lowest accuracy typically occurs on Sundays and around 5 am. However, apart from these times, the accuracy scores tend to be relatively high, which means that both datasets reflect mobility changes in a similar way.

\section*{Discussion and Conclusions}\label{sec:discussion}

Natural disasters, such as wildfires, have a significant impact on communities in affected areas, with some vulnerable populations more exposed to the consequences of the emergency than others~\cite{winker_wildfires_2024, chas-amil_spatial_2022}. In this paper, we show how wildfires disproportionately affect poorer populations in the Valparaíso region of Chile. We show that individuals from the lower socioeconomic strata left their homes with a one-day time lag and remained displaced for a longer period of time, with their travelled distances decreasing during this time. Furthermore, we identify distinct patterns of mobility segregation among evacuated populations, with irregular displacement patterns observed on the first night of the wildfire, followed by more structured movement toward areas of similar socioeconomic status for at least five subsequent nights.

One of the key contributions of our work is the comparative analysis of human mobility using both mobile phone data and Facebook's crisis mobility data provided by Meta. This comparison is important because while mobile phone data offers real-time, continuous monitoring of individuals' movements during the event, the Facebook dataset provides broader population-level insights, albeit with higher temporal and spatial aggregation. Our results show that, while the datasets are inherently different, they exhibit some degree of comparability in recording activity changes over the overlapping period. This finding is significant because it highlights the potential for combining both data sources to create a more comprehensive understanding of displacement during natural disasters.

One limitation in this article is that we analyse the consequences of one single natural disaster: the Valparaíso wildfire. However, the available literature on the impacts of natural disasters in general and wildfires in particular similarly indicates their uneven impact on people of different socioeconomic backgrounds and the subsequent increase in economic inequality~\cite{yabe_effects_2020, deng_high-resolution_2021, winker_wildfires_2024, lindersson_wider_2023, jayash_paudel_natural_2023}. Therefore, we expect similar behavioural patterns to occur in other natural disasters as well.

Other limitations include our choice of heuristics for identifying users' home locations and their socioeconomic status based on census zones. As in the previous limitation, the methods employed are consistent with current research using communication data~\cite{pappalardo_evaluation_2021, elejalde_social_2024}. Furthermore, the alignment of our results with the findings of previous studies supports the validity and robustness of our methodology.

Future research should focus on differentiating the impact of various types of natural disasters on human mobility and displacement patterns. This is important because different disasters, such as wildfires, floods, and earthquakes, can trigger distinct displacement dynamics due to varying levels of severity, duration, and geographic scope. Understanding these nuances would provide deeper insight into how people respond to specific types of crises, allowing more targeted and effective disaster preparedness and response strategies. In addition, it would help policy makers and humanitarian organisations allocate resources more efficiently based on the nature of the disaster.

Establishing long-term relationships with industry partners is essential for continuous access to data, especially in the context of disaster management. When natural disasters or similar emergencies occur, timely access to data can facilitate the implementation of disaster response plans and enable researchers to develop strategies that ensure equitable access to necessary assistance and resources for those in greatest need.

\section*{Acknowledgements}
\label{sec:acks}

L.F. thanks the funding and support of Telefónica Chile. This research was supported by FONDECYT Grant N°1221315 to Leo Ferres. L.F. also acknowledges financial support from the Lagrange Project of the Institute for Scientific Interchange Foundation (ISI Foundation), funded by Fondazione Cassa di Risparmio di Torino (Fondazione CRT). M.K. acknowledges funding from the National Laboratory for Health Security (RRF-2.3.1-21-2022-00006), the ANR project DATAREDUX (ANR-19-CE46-0008); the SoBigData++ H2020-871042; and the MOMA WWTF project.

\clearpage
\printbibliography

\clearpage

\setcounter{figure}{0} 
\captionsetup[figure]{name=Supplementary Figure}
\setcounter{table}{0} 
\captionsetup[table]{name=Supplementary Table}
\section*{Supplementary Information}


\subsection{Comparison of Population}

\begin{figure}[H]
    \centering
    \includegraphics[width=0.5\linewidth]{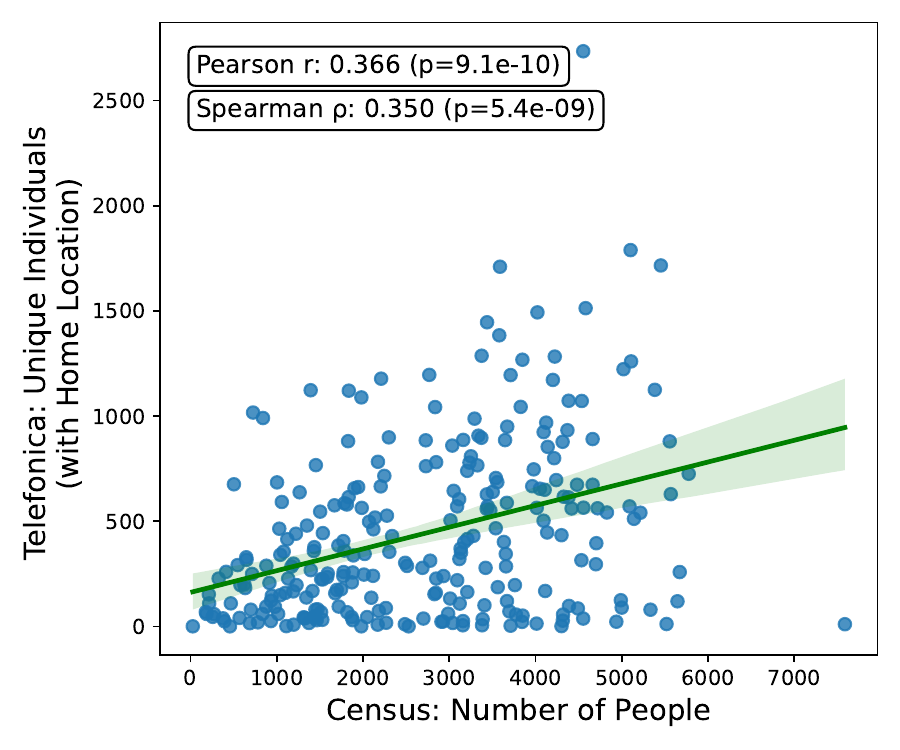}
    \caption{
        \textbf{Correlations between census zone populations as represented in Telefónica data and official statistics.}
        }
    \label{fig:plot11_1}
\end{figure}

\begin{figure}[H]
    \centering
    \includegraphics[width=1\linewidth]{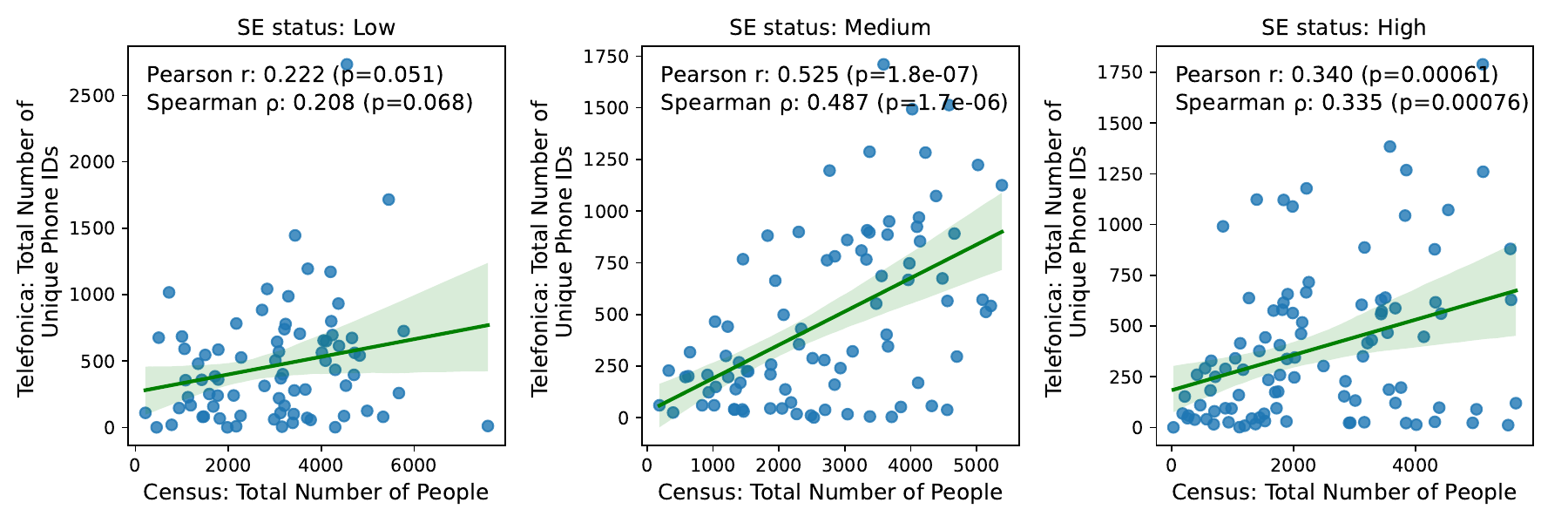}
    \caption{
            \textbf{Correlations between census zone populations per each socioeconomic status as represented in Telefónica data and official statistics.} The Medium SE class is better represented in Telefónica data, while the Low SE class is the most underrepresented (when compared with the official statistics).
            }
    \label{fig:plot11_2}
\end{figure}

\clearpage
\subsection{Distribution of Towers}

\begin{figure}[H]
    \centering
    \includegraphics[width=1\linewidth]{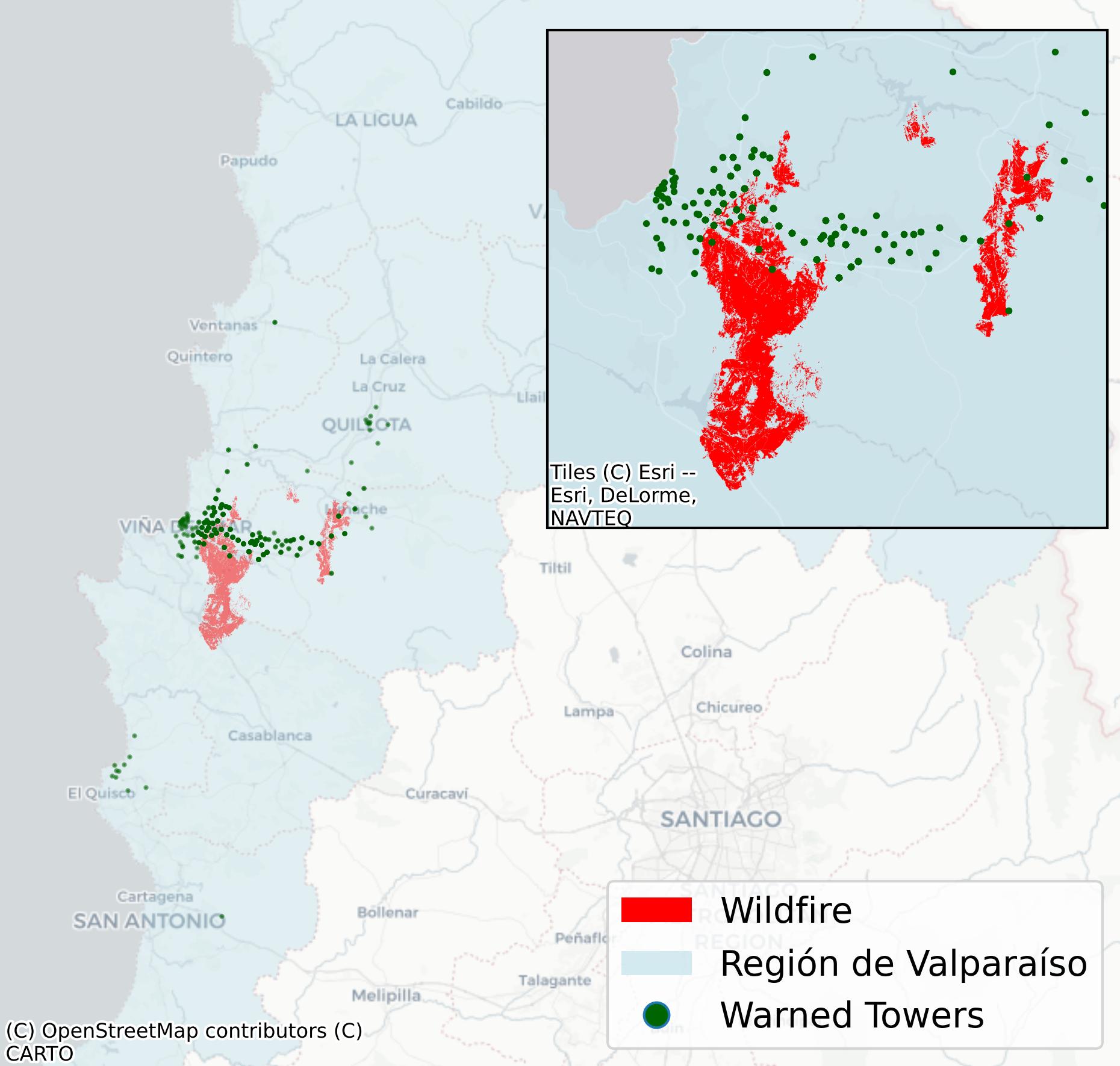}
    \caption{
        \textbf{Map of the areas affected by the wildfire and spatial distribution of warned towers.} 
        }
    \label{fig:plot12}
\end{figure}

\clearpage
\subsection{Assortativity}

\begin{figure}[H]
    \centering
    \begin{adjustbox}{width=1\textwidth,center}
        \includegraphics{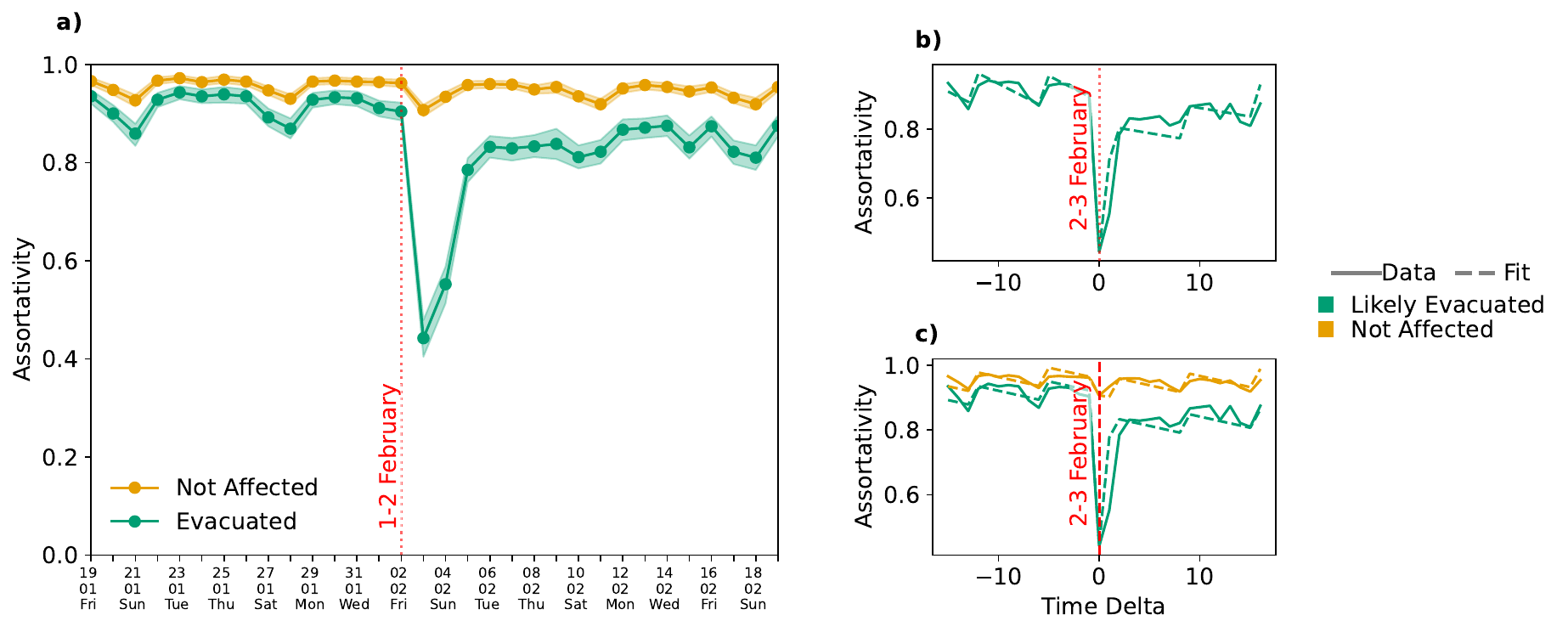}
    \end{adjustbox}
    \caption{
        \textbf{Assortativity of all people (moved and stayed).} \textbf{a)} Assortativity over time for two groups of people (Not Affected and Likely Evacuated), with a 95\% confidence interval.  The 95\% CI was obtained through bootstrapping the assortativity values of all people (moved and stayed) by iteratively resampling the original dataset (1,000 times) with replacement. For each resample, an assortativity measure was computed using a heatmap representing movement patterns between socioeconomic categories, capturing the variability in assortativity over time. \textbf{b)} Assortativity of likely evacuated people as given in data and inferred with the regression discontinuity in time model. \textbf{c)} Assortativity of likely evacuated and not affected people as given in data and inferred with the difference-in-differences model.
        }
    \label{fig:plot5_all}
\end{figure}

\clearpage
\subsection{Comparison of Telefónica and Meta: Z-Scores}

\begin{figure}[H]
    \centering
    \begin{adjustbox}{width=1\textwidth,center}
        \includegraphics{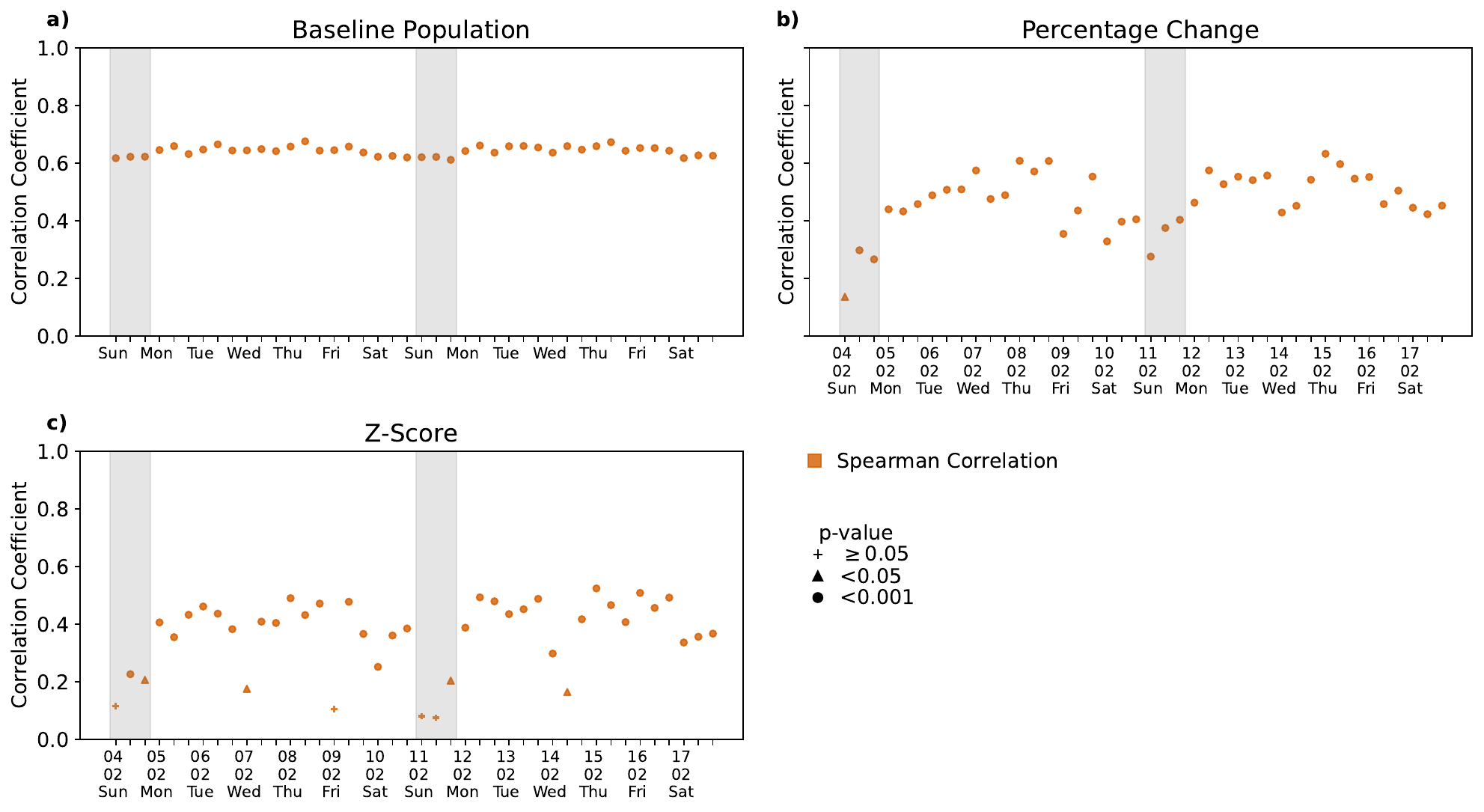}
    \end{adjustbox}
    \caption{
            \textbf{Spearman correlations between three indicators in Telefónica and Meta datasets across geolocated cells over time.} \textbf{a)} Comparing recorded number of users between two datasets in pre-crisis period over time. Sundays coloured grey.  \textbf{b)} Comparing percentage changes of users activity before and after the crisis between two datasets over time. Sundays coloured grey.  \textbf{c)} Comparing changes in z-scores of users activity before and after the crisis between two datasets over time. Sundays coloured grey. }
    \label{fig:plot13_1}
\end{figure}

\begin{figure}[H]
    \centering
    \begin{adjustbox}{width=0.7\textwidth,center}
    \includegraphics{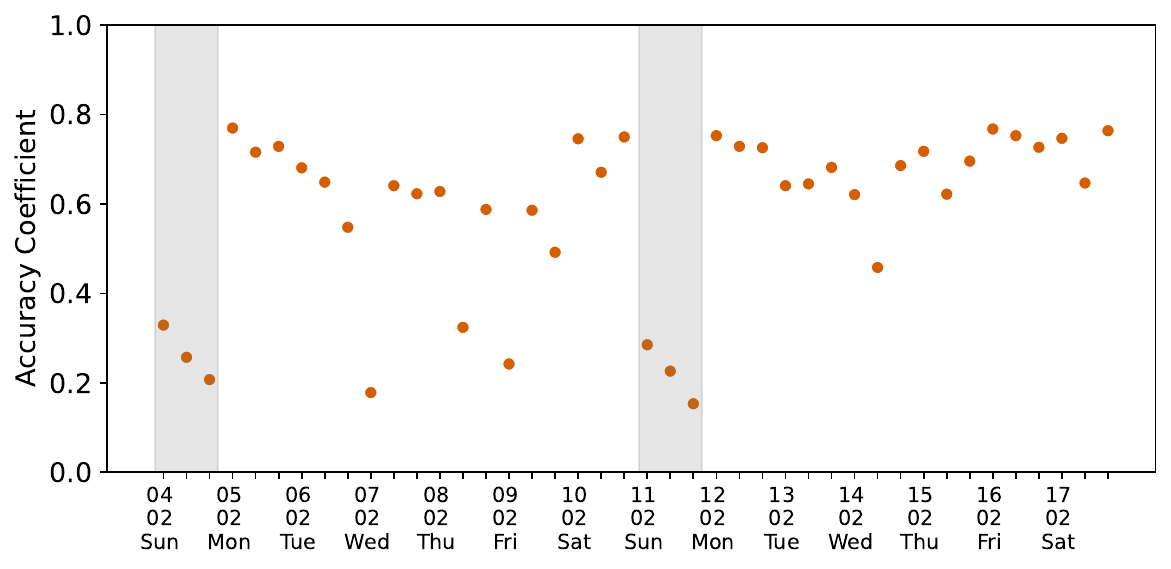}
    \end{adjustbox}
    \caption{
        \textbf{Accuracy scores for the Z-Score in different time periods.} Sundays coloured grey.
        }
    \label{fig:plot13_2}
\end{figure}

\clearpage

\subsection{Regression Tables}

\begin {table}[H]
\begin{center}
\begin{threeparttable}
    \caption{RDiT Regression Results for the Fraction of People}
    \setlength\extrarowheight{-2pt}
    \begin{tabular}{lcccccc}
    \toprule
     & Model 1 & Model 2 & Model 3 & Model 4 & Model 5 & Model 6 \\
    \multicolumn{7}{c}{Dependent Variable: Fraction of People} \\
    \midrule
    Intercept & 0.067*** & 0.164*** & 0.096*** & 0.074*** & 0.168*** & 0.102*** \\
      & (0.020) & (0.035) & (0.021) & (0.014) & (0.027) & (0.015) \\
    Threshold & 0.267*** & 0.258*** & 0.351*** & 0.336*** & 0.328*** & 0.417*** \\
      & (0.064) & (0.054) & (0.051) & (0.038) & (0.031) & (0.034) \\
    Time Delta & 0.001 & 0.033*** & 0.014* & 0.001 & 0.032*** & 0.013** \\
      & (0.002) & (0.009) & (0.006) & (0.001) & (0.007) & (0.004) \\
    Time Delta:Threshold & -0.018** & -0.082*** & -0.069*** & -0.018*** & -0.080*** & -0.067*** \\
      & (0.006) & (0.019) & (0.013) & (0.004) & (0.015) & (0.011) \\
    Weekday & 0.021*** & 0.019*** & 0.020*** & 0.021*** & 0.018*** & 0.019*** \\
      & (0.006) & (0.004) & (0.004) & (0.004) & (0.003) & (0.003) \\
    High SE class &   &   &   & -0.010 & -0.010 & -0.010 \\
      &   &   &   & (0.009) & (0.013) & (0.008) \\
    High SE class:Threshold &   &   &   & -0.098*** & -0.098*** & -0.098*** \\
      &   &   &   & (0.028) & (0.027) & (0.025) \\
    Medium SE class &   &   &   & -0.007 & -0.007 & -0.007 \\
      &   &   &   & (0.007) & (0.011) & (0.007) \\
    Medium SE class:Threshold &   &   &   & -0.070** & -0.070** & -0.070** \\
      &   &   &   & (0.026) & (0.024) & (0.023) \\
    Time Delta ** 2 &   & 0.002*** & 0.002*** &   & 0.002*** & 0.002*** \\
      &   & (0.001) & ($<$0.001) &   & ($<$0.001) & ($<$0.001) \\
    Time Delta ** 3 &   &   & $<$0.001*** &   &   & $<$0.001*** \\
      &   &   & ($<$0.001) &   &   & ($<$0.001) \\
    \midrule
    R-squared & 0.750 & 0.852 & 0.894 & 0.738 & 0.816 & 0.850 \\
    Adjusted R-squared & 0.714 & 0.823 & 0.869 & 0.714 & 0.797 & 0.832 \\
    No. observations & 32 & 32 & 32 & 96 & 96 & 96 \\
    \bottomrule
    \end{tabular}
    \begin{tablenotes}
    \item \textit{Notes:} Standard Errors are in parentheses. Standard Errors are heteroscedasticity and autocorrelation robust (HAC) using 1 lag. The baseline for the SE status is Low. \newline
    *** p$<$0.001, ** p$<$0.01, * p$<$0.05.
    \end{tablenotes}
    \label{tab:regression_results_fraction_rdd}
\end{threeparttable}
\end{center}
\end{table}

\begin{tabular}{lcccccc}

\bottomrule
\end{tabular}

\begin {table}[H]
\begin{center}
\begin{threeparttable}
    \caption{DiD Regression Results for the Fraction of People}
    \setlength\extrarowheight{-2pt}
    \begin{tabular}{lcccccc}
    \toprule
     & Model 1 & Model 2 & Model 3 & Model 4 & Model 5 & Model 6 \\
    \multicolumn{7}{c}{Dependent Variable: Fraction of People} \\
    \midrule
    Intercept & 0.006 & 0.024 & -0.015 & 0.005 & 0.005 & -0.014 \\
      & (0.033) & (0.022) & (0.037) & (0.021) & (0.021) & (0.025) \\
    Threshold & 0.103* & 0.094* & 0.169* & 0.106*** & 0.106*** & 0.170*** \\
      & (0.043) & (0.037) & (0.068) & (0.029) & (0.029) & (0.046) \\
    Treatment:Threshold & 0.118*** & 0.118*** & 0.118*** & 0.181*** & 0.181*** & 0.181*** \\
      & (0.034) & (0.033) & (0.031) & (0.030) & (0.030) & (0.027) \\
    Time Delta & -0.005 & -0.004 & -0.015* & -0.005** & -0.005** & -0.015** \\
      & (0.003) & (0.002) & (0.007) & (0.002) & (0.002) & (0.005) \\
    Treatment & 0.031* & 0.031** & 0.031* & 0.038** & 0.038** & 0.038*** \\
      & (0.014) & (0.010) & (0.014) & (0.012) & (0.012) & (0.011) \\
    Weekday & 0.015*** & 0.016*** & 0.017*** & 0.015*** & 0.015*** & 0.016*** \\
      & (0.004) & (0.004) & (0.004) & (0.003) & (0.003) & (0.003) \\
    High SE class &   &   &   & 0.004 & 0.004 & 0.004 \\
      &   &   &   & (0.009) & (0.009) & (0.008) \\
    High SE class:Threshold &   &   &   & -0.002 & -0.002 & -0.002 \\
      &   &   &   & (0.013) & (0.013) & (0.014) \\
    High SE class:Treatment &   &   &   & -0.014 & -0.014 & -0.014 \\
      &   &   &   & (0.014) & (0.014) & (0.012) \\
    High SE class:Treat:Thres &   &   &   & -0.097** & -0.097** & -0.097** \\
      &   &   &   & (0.037) & (0.037) & (0.034) \\
    Medium SE class &   &   &   & -0.003 & -0.003 & -0.003 \\
      &   &   &   & (0.008) & (0.008) & (0.007) \\
    Medium SE class:Threshold &   &   &   & 0.002 & 0.002 & 0.002 \\
      &   &   &   & (0.012) & (0.012) & (0.013) \\
    Medium SE class:Treatment &   &   &   & -0.005 & -0.005 & -0.005 \\
      &   &   &   & (0.012) & (0.012) & (0.012) \\
    Medium SE class:Treat:Thres &   &   &   & -0.072* & -0.072* & -0.072* \\
      &   &   &   & (0.033) & (0.033) & (0.031) \\
    Time Delta ** 2 &   & -$<$0.001 & -$<$0.001 &   &   & -$<$0.001** \\
      &   & ($<$0.001) & ($<$0.001) &   &   & ($<$0.001) \\
    Time Delta ** 3 &   &   & $<$0.001* &   &   & $<$0.001** \\
      &   &   & ($<$0.001) &   &   & ($<$0.001) \\
    \midrule
    R-squared & 0.666 & 0.685 & 0.733 & 0.663 & 0.663 & 0.721 \\
    Adjusted R-squared & 0.637 & 0.652 & 0.699 & 0.639 & 0.639 & 0.697 \\
    No. observations & 64 & 64 & 64 & 192 & 192 & 192 \\
    \bottomrule
    \end{tabular}
    \begin{tablenotes}
    \item \textit{Notes:} Standard Errors are in parentheses. Standard Errors are heteroscedasticity and autocorrelation robust (HAC) using 1 lag. The baseline for the SE status is Low. \newline
    *** p$<$0.001, ** p$<$0.01, * p$<$0.05.
    \end{tablenotes}
    \label{tab:regression_results_fraction_dd}
\end{threeparttable}
\end{center}
\end{table}

\begin {table}[H]
\begin{center}
\begin{threeparttable}
    \caption{RDiT Regression Results for the Mean kilometres Moved}
    \setlength\extrarowheight{-2pt}
    \begin{tabular}{lcccccc}
    \toprule
     & Model 1 & Model 2 & Model 3 & Model 4 & Model 5 & Model 6 \\
    \multicolumn{7}{c}{Dependent Variable: Mean kilometres Moved} \\
    \midrule
    Intercept & 11.274*** & 10.953*** & 11.628*** & 10.318*** & 10.008*** & 10.682*** \\
      & (0.527) & (0.571) & (0.643) & (0.472) & (0.572) & (0.618) \\
    Threshold & -1.543* & -1.609** & -2.395*** & -1.841*** & -1.904*** & -2.689*** \\
      & (0.616) & (0.565) & (0.661) & (0.477) & (0.485) & (0.659) \\
    Time Delta & -0.083* & -0.239* & -0.054 & -0.084* & -0.234* & -0.050 \\
      & (0.037) & (0.093) & (0.123) & (0.035) & (0.118) & (0.121) \\
    Time Delta:Threshold & 0.239*** & 0.569*** & 0.417* & 0.240*** & 0.558* & 0.406* \\
      & (0.049) & (0.170) & (0.172) & (0.051) & (0.231) & (0.205) \\
    High SE class &   &   &   & 1.273*** & 1.273*** & 1.273*** \\
      &   &   &   & (0.342) & (0.343) & (0.333) \\
    High SE class:Threshold &   &   &   & 0.114 & 0.114 & 0.114 \\
      &   &   &   & (0.540) & (0.538) & (0.530) \\
    Medium SE class &   &   &   & 1.366*** & 1.366*** & 1.366*** \\
      &   &   &   & (0.224) & (0.236) & (0.211) \\
    Medium SE class:Threshold &   &   &   & 1.087* & 1.087* & 1.087* \\
      &   &   &   & (0.493) & (0.499) & (0.490) \\
    Tuesday & -2.693*** & -2.821*** & -2.895*** & -2.783*** & -2.905*** & -2.979*** \\
      & (0.546) & (0.521) & (0.522) & (0.432) & (0.415) & (0.413) \\
    Wednesday & -3.653*** & -3.790*** & -3.879*** & -3.669*** & -3.800*** & -3.889*** \\
      & (0.416) & (0.374) & (0.377) & (0.455) & (0.433) & (0.438) \\
    Thursday & -3.649*** & -3.772*** & -3.887*** & -3.690*** & -3.809*** & -3.923*** \\
      & (0.677) & (0.623) & (0.610) & (0.563) & (0.526) & (0.518) \\
    Friday & -2.931*** & -3.020*** & -3.165*** & -2.939*** & -3.025*** & -3.170*** \\
      & (0.485) & (0.493) & (0.412) & (0.417) & (0.427) & (0.384) \\
    Saturday & -1.583*** & -1.543*** & -1.636*** & -1.590*** & -1.551*** & -1.645*** \\
      & (0.475) & (0.414) & (0.412) & (0.438) & (0.405) & (0.398) \\
    Sunday & -1.277*** & -1.268*** & -1.311*** & -1.274** & -1.265** & -1.308** \\
      & (0.361) & (0.341) & (0.342) & (0.443) & (0.421) & (0.415) \\
    Time Delta ** 2 &   & -0.011* & -0.005 &   & -0.010 & -0.005 \\
      &   & (0.005) & (0.005) &   & (0.007) & (0.006) \\
    Time Delta ** 3 &   &   & -0.001* &   &   & -0.001* \\
      &   &   & ($<$0.001) &   &   & ($<$0.001) \\
    \midrule
    R-squared & 0.842 & 0.856 & 0.870 & 0.712 & 0.720 & 0.729 \\
    Adjusted R-squared & 0.775 & 0.783 & 0.794 & 0.665 & 0.670 & 0.676 \\
    No. observations & 31 & 31 & 31 & 93 & 93 & 93 \\
    \bottomrule
    \end{tabular}
    \begin{tablenotes}
    \item \textit{Notes:} Standard Errors are in parentheses. Standard Errors are heteroscedasticity and autocorrelation robust (HAC) using 1 lag. The baselines for the weekday is Monday, and for the SE status is Low. \newline
    *** p$<$0.001, ** p$<$0.01, * p$<$0.05.
    \end{tablenotes}
    \label{tab:regression_results_mean_rdd}
\end{threeparttable}
\end{center}
\end{table}

\begin {table}[H]
\begin{center}
\begin{threeparttable}
    \caption{DiD Regression Results for the Mean kilometres Moved}
    \setlength\extrarowheight{-2pt}
    \begin{tabular}{lcccccc}
    \toprule
     & Model 1 & Model 2 & Model 3 & Model 4 & Model 5 & Model 6 \\
    \multicolumn{7}{c}{Dependent Variable: Mean kilometres Moved} \\
    \midrule
    Intercept & 14.855*** & 14.805*** & 14.624*** & 15.332*** & 15.286*** & 15.107*** \\
      & (0.360) & (0.419) & (0.420) & (0.370) & (0.386) & (0.385) \\
    Threshold & -1.071 & -0.968 & -1.560** & -1.445** & -1.350* & -1.942*** \\
      & (0.579) & (0.704) & (0.532) & (0.534) & (0.580) & (0.554) \\
    Treatment:Threshold & -1.205* & -1.413 & -1.413 & -1.142* & -1.332 & -1.332 \\
      & (0.473) & (1.238) & (0.798) & (0.524) & (0.865) & (0.705) \\
    Time Delta & 0.084** & 0.077* & 0.193*** & 0.084*** & 0.078** & 0.194*** \\
      & (0.031) & (0.038) & (0.042) & (0.024) & (0.027) & (0.045) \\
    Time Delta:Treatment &   & 0.013 & 0.013 &   & 0.012 & 0.012 \\
      &   & (0.063) & (0.036) &   & (0.046) & (0.034) \\
    Treatment & -1.766*** & -1.666** & -1.666*** & -3.207*** & -3.115*** & -3.115*** \\
      & (0.332) & (0.563) & (0.388) & (0.352) & (0.459) & (0.355) \\
    High SE class &   &   &   & -0.984** & -0.984** & -0.984*** \\
      &   &   &   & (0.311) & (0.308) & (0.278) \\
    High SE class:Threshold &   &   &   & 0.536 & 0.536 & 0.536 \\
      &   &   &   & (0.481) & (0.480) & (0.458) \\
    High SE class:Treatment &   &   &   & 2.257*** & 2.257*** & 2.257*** \\
      &   &   &   & (0.482) & (0.482) & (0.447) \\
    High SE class:Treat:Thres &   &   &   & -0.422 & -0.422 & -0.422 \\
      &   &   &   & (0.767) & (0.767) & (0.732) \\
    Medium SE class &   &   &   & -0.348 & -0.348 & -0.348 \\
      &   &   &   & (0.366) & (0.363) & (0.319) \\
    Medium SE class:Threshold &   &   &   & 0.572 & 0.572 & 0.572 \\
      &   &   &   & (0.506) & (0.505) & (0.458) \\
    Medium SE class:Treatment &   &   &   & 1.714*** & 1.714*** & 1.714*** \\
      &   &   &   & (0.450) & (0.450) & (0.381) \\
    Medium SE class:Treat:Thres &   &   &   & 0.515 & 0.515 & 0.515 \\
      &   &   &   & (0.733) & (0.733) & (0.674) \\
    Tuesday & -2.959*** & -2.959*** & -2.921*** & -2.984*** & -2.984*** & -2.947*** \\
      & (0.386) & (0.386) & (0.345) & (0.350) & (0.350) & (0.301) \\
    Wednesday & -4.569*** & -4.569*** & -4.519*** & -4.564*** & -4.564*** & -4.515*** \\
      & (0.355) & (0.354) & (0.297) & (0.338) & (0.338) & (0.310) \\
    Thursday & -4.810*** & -4.810*** & -4.760*** & -4.835*** & -4.835*** & -4.786*** \\
      & (0.455) & (0.456) & (0.407) & (0.382) & (0.383) & (0.351) \\
    Friday & -3.899*** & -3.899*** & -3.857*** & -3.900*** & -3.900*** & -3.859*** \\
      & (0.320) & (0.320) & (0.273) & (0.289) & (0.289) & (0.262) \\
    Saturday & -1.475*** & -1.475*** & -1.629*** & -1.470*** & -1.470*** & -1.624*** \\
      & (0.435) & (0.431) & (0.328) & (0.372) & (0.370) & (0.298) \\
    Sunday & -2.018*** & -2.018*** & -2.090*** & -2.007*** & -2.007*** & -2.079*** \\
      & (0.340) & (0.340) & (0.293) & (0.347) & (0.348) & (0.295) \\
    Time Delta ** 2 &   &   & 0.007*** &   &   & 0.007*** \\
      &   &   & (0.001) &   &   & (0.001) \\
    Time Delta ** 3 &   &   & -0.001*** &   &   & -0.001*** \\
      &   &   & ($<$0.001) &   &   & ($<$0.001) \\
    \midrule
    R-squared & 0.865 & 0.865 & 0.915 & 0.785 & 0.785 & 0.825 \\
    Adjusted R-squared & 0.838 & 0.835 & 0.892 & 0.762 & 0.761 & 0.803 \\
    No. observations & 62 & 62 & 62 & 186 & 186 & 186 \\
    \bottomrule
    \end{tabular}
    \begin{tablenotes}
    \item \textit{Notes:} Standard Errors are in parentheses. Standard Errors are heteroscedasticity and autocorrelation robust (HAC) using 1 lag. The baselines for the weekday is Monday, and for the SE status is Low. \newline
    *** p$<$0.001, ** p$<$0.01, * p$<$0.05.
    \end{tablenotes}
    \label{tab:regression_results_mean_dd}
\end{threeparttable}
\end{center}
\end{table}

\begin {table}[H]
\begin{center}
\begin{threeparttable}
    \caption{RDiT Regression Results for the Median kilometres Moved}
    \setlength\extrarowheight{-2pt}
    \begin{tabular}{lcccccc}
    \toprule
     & Model 1 & Model 2 & Model 3 & Model 4 & Model 5 & Model 6 \\
    \multicolumn{7}{c}{Dependent Variable: Median kilometres Moved} \\
    \midrule
    Intercept & 4.013*** & 4.000*** & 4.258*** & 3.938*** & 3.928*** & 4.135*** \\
      & (0.171) & (0.172) & (0.161) & (0.149) & (0.157) & (0.155) \\
    Threshold & -0.418* & -0.420* & -0.722*** & -0.596*** & -0.598*** & -0.839*** \\
      & (0.171) & (0.173) & (0.206) & (0.155) & (0.156) & (0.191) \\
    Time Delta & -0.021 & -0.027 & 0.044 & -0.022 & -0.027 & 0.030 \\
      & (0.018) & (0.039) & (0.051) & (0.012) & (0.032) & (0.036) \\
    Time Delta:Threshold & 0.039 & 0.053 & -0.005 & 0.047** & 0.057 & 0.011 \\
      & (0.022) & (0.084) & (0.091) & (0.017) & (0.069) & (0.069) \\
    High SE class &   &   &   & -0.190 & -0.190 & -0.190 \\
      &   &   &   & (0.115) & (0.115) & (0.113) \\
    High SE class:Threshold &   &   &   & 0.127 & 0.127 & 0.127 \\
      &   &   &   & (0.167) & (0.167) & (0.165) \\
    Medium SE class &   &   &   & 0.339*** & 0.339*** & 0.339*** \\
      &   &   &   & (0.083) & (0.084) & (0.078) \\
    Medium SE class:Threshold &   &   &   & 0.403* & 0.403* & 0.403* \\
      &   &   &   & (0.162) & (0.162) & (0.159) \\
    Tuesday & -1.359*** & -1.364*** & -1.393*** & -1.336*** & -1.340*** & -1.363*** \\
      & (0.129) & (0.130) & (0.125) & (0.127) & (0.129) & (0.128) \\
    Wednesday & -1.450*** & -1.456*** & -1.490*** & -1.501*** & -1.506*** & -1.533*** \\
      & (0.156) & (0.156) & (0.141) & (0.146) & (0.148) & (0.143) \\
    Thursday & -1.395*** & -1.401*** & -1.444*** & -1.378*** & -1.382*** & -1.417*** \\
      & (0.137) & (0.137) & (0.123) & (0.139) & (0.140) & (0.136) \\
    Friday & -1.275*** & -1.279*** & -1.335*** & -1.221*** & -1.224*** & -1.269*** \\
      & (0.109) & (0.115) & (0.092) & (0.118) & (0.120) & (0.114) \\
    Saturday & -0.749*** & -0.747*** & -0.783*** & -0.739*** & -0.738*** & -0.767*** \\
      & (0.138) & (0.142) & (0.157) & (0.140) & (0.142) & (0.149) \\
    Sunday & -0.326* & -0.326* & -0.342* & -0.317* & -0.316* & -0.329* \\
      & (0.146) & (0.145) & (0.140) & (0.161) & (0.160) & (0.158) \\
    Time Delta ** 2 &   & -$<$0.001 & 0.002 &   & -$<$0.001 & 0.001 \\
      &   & (0.003) & (0.003) &   & (0.002) & (0.002) \\
    Time Delta ** 3 &   &   & -$<$0.001* &   &   & -$<$0.001 \\
      &   &   & ($<$0.001) &   &   & ($<$0.001) \\
    \midrule
    R-squared & 0.871 & 0.871 & 0.885 & 0.805 & 0.805 & 0.811 \\
    Adjusted R-squared & 0.816 & 0.807 & 0.819 & 0.773 & 0.770 & 0.775 \\
    No. observations & 31 & 31 & 31 & 93 & 93 & 93 \\
    \bottomrule
    \end{tabular}
    \begin{tablenotes}
    \item \textit{Notes:} Standard Errors are in parentheses. Standard Errors are heteroscedasticity and autocorrelation robust (HAC) using 1 lag. The baselines for the weekday is Monday, and for the SE status is Low. \newline
    *** p$<$0.001, ** p$<$0.01, * p$<$0.05.
    \end{tablenotes}
    \label{tab:regression_results_median_rdd}
\end{threeparttable}
\end{center}
\end{table}

\begin {table}[H]
\begin{center}
\begin{threeparttable}
    \caption{DiD Regression Results for the Median kilometres Moved}
    \setlength\extrarowheight{-2pt}
    \begin{tabular}{lcccccc}
    \toprule
     & Model 1 & Model 2 & Model 3 & Model 4 & Model 5 & Model 6 \\
    \multicolumn{7}{c}{Dependent Variable: Median kilometres Moved} \\
    \midrule
    Intercept & 4.323*** & 4.329*** & 4.230*** & 4.877*** & 4.818*** & 4.667*** \\
      & (0.118) & (0.116) & (0.134) & (0.136) & (0.154) & (0.187) \\
    Threshold & -0.004 & -0.015 & -0.097 & -0.059 & 0.063 & 0.085 \\
      & (0.175) & (0.222) & (0.229) & (0.228) & (0.283) & (0.344) \\
    Treatment:Threshold & -0.596*** & -0.574 & -0.574* & -0.614** & -0.857* & -0.857** \\
      & (0.151) & (0.301) & (0.241) & (0.237) & (0.335) & (0.327) \\
    Time Delta & 0.010 & 0.011 & 0.031 & 0.006 & -0.002 & 0.003 \\
      & (0.009) & (0.013) & (0.019) & (0.009) & (0.015) & (0.031) \\
    Time Delta:Treatment &   & -0.001 & -0.001 &   & 0.016 & 0.016 \\
      &   & (0.018) & (0.013) &   & (0.018) & (0.017) \\
    Treatment & -0.042 & -0.053 & -0.053 & -0.728*** & -0.610*** & -0.610*** \\
      & (0.104) & (0.147) & (0.122) & (0.116) & (0.163) & (0.160) \\
    High SE class &   &   &   & -1.153*** & -1.153*** & -1.153*** \\
      &   &   &   & (0.114) & (0.110) & (0.101) \\
    High SE class:Threshold &   &   &   & 0.028 & 0.028 & 0.028 \\
      &   &   &   & (0.216) & (0.217) & (0.202) \\
    High SE class:Treatment &   &   &   & 0.963*** & 0.963*** & 0.963*** \\
      &   &   &   & (0.154) & (0.151) & (0.148) \\
    High SE class:Treat:Thres &   &   &   & 0.099 & 0.099 & 0.099 \\
      &   &   &   & (0.275) & (0.276) & (0.265) \\
    Medium SE class &   &   &   & -0.857*** & -0.858*** & -0.857*** \\
      &   &   &   & (0.124) & (0.121) & (0.114) \\
    Medium SE class:Threshold &   &   &   & 0.592 & 0.592 & 0.592 \\
      &   &   &   & (0.367) & (0.366) & (0.364) \\
    Medium SE class:Treatment &   &   &   & 1.196*** & 1.196*** & 1.196*** \\
      &   &   &   & (0.146) & (0.146) & (0.135) \\
    Medium SE class:Treat:Thres &   &   &   & -0.189 & -0.189 & -0.189 \\
      &   &   &   & (0.404) & (0.403) & (0.399) \\
    Tuesday & -1.427*** & -1.427*** & -1.404*** & -1.349*** & -1.349*** & -1.313*** \\
      & (0.120) & (0.120) & (0.110) & (0.102) & (0.102) & (0.098) \\
    Wednesday & -1.577*** & -1.577*** & -1.549*** & -1.571*** & -1.571*** & -1.528*** \\
      & (0.130) & (0.130) & (0.107) & (0.110) & (0.111) & (0.106) \\
    Thursday & -1.501*** & -1.501*** & -1.471*** & -1.459*** & -1.459*** & -1.412*** \\
      & (0.127) & (0.127) & (0.107) & (0.117) & (0.119) & (0.116) \\
    Friday & -1.427*** & -1.427*** & -1.399*** & -1.362*** & -1.362*** & -1.314*** \\
      & (0.108) & (0.108) & (0.094) & (0.101) & (0.101) & (0.100) \\
    Saturday & -0.632*** & -0.632*** & -0.666*** & -0.569*** & -0.569*** & -0.591*** \\
      & (0.127) & (0.127) & (0.116) & (0.132) & (0.131) & (0.134) \\
    Sunday & -0.250 & -0.250 & -0.266* & -0.195 & -0.195 & -0.205 \\
      & (0.128) & (0.128) & (0.115) & (0.134) & (0.133) & (0.127) \\
    Time Delta ** 2 &   &   & 0.002*** &   &   & 0.002** \\
      &   &   & ($<$0.001) &   &   & (0.001) \\
    Time Delta ** 3 &   &   & -$<$0.001* &   &   & -$<$0.001 \\
      &   &   & ($<$0.001) &   &   & ($<$0.001) \\
    \midrule
    R-squared & 0.871 & 0.871 & 0.902 & 0.717 & 0.719 & 0.732 \\
    Adjusted R-squared & 0.845 & 0.842 & 0.875 & 0.687 & 0.687 & 0.698 \\
    No. observations & 62 & 62 & 62 & 186 & 186 & 186 \\
    \bottomrule
    \end{tabular}
    \begin{tablenotes}
    \item \textit{Notes:} Standard Errors are in parentheses. Standard Errors are heteroscedasticity and autocorrelation robust (HAC) using 1 lag. The baselines for the weekday is Monday, and for the SE status is Low. \newline
    *** p$<$0.001, ** p$<$0.01, * p$<$0.05.
    \end{tablenotes}
    \label{tab:regression_results_median_dd}
\end{threeparttable}
\end{center}
\end{table}

\begin {table}[H]
\begin{center}
\begin{threeparttable}
    \caption{RDiT Regression Results for the Assortativity (Moved)}
    \setlength\extrarowheight{-2pt}
    \begin{tabular}{lccc}
    \toprule
     & Model 1 & Model 2 & Model 3 \\
    \multicolumn{4}{c}{Dependent Variable: Assortativity} \\
    \midrule
     Intercept & 0.233*** & 0.247*** & 0.273*** \\
      & (0.021) & (0.025) & (0.018) \\
    First Night (2-3 February) & -0.236*** & -0.251*** & -0.232*** \\
      & (0.025) & (0.035) & (0.034) \\
    Threshold & 0.108*** & 0.110*** & 0.067* \\
      & (0.030) & (0.029) & (0.031) \\
    Time Delta & -0.003 & 0.002 & 0.009 \\
      & (0.002) & (0.007) & (0.005) \\
    Time Delta:Threshold & -0.004 & -0.014 & -0.017 \\
      & (0.003) & (0.015) & (0.012) \\
    Weekday & -0.008* & -0.008** & -0.009** \\
      & (0.003) & (0.003) & (0.003) \\
    Time Delta ** 2 &   & $<$0.001 & $<$0.001 \\
      &   & ($<$0.001) & ($<$0.001) \\
    Time Delta ** 3 &   &   & -$<$0.001 \\
      &   &   & ($<$0.001) \\
    \midrule
    R-squared & 0.604 & 0.612 & 0.642 \\
    Adjusted R-squared & 0.528 & 0.519 & 0.538 \\
    No. observations & 32 & 32 & 32 \\
    \bottomrule
    \end{tabular}
    \begin{tablenotes}
    \item \textit{Notes:} Standard Errors are in parentheses. Standard Errors are heteroscedasticity and autocorrelation robust (HAC) using 1 lag. \newline
    *** p$<$0.001, ** p$<$0.01, * p$<$0.05.
    \end{tablenotes}
    \label{tab:regression_results_assort_moved_rdd}
\end{threeparttable}
\end{center}
\end{table}

\begin {table}[H]
\begin{center}
\begin{threeparttable}
    \caption{DiD Regression Results for the Assortativity (Moved)}
    \setlength\extrarowheight{-2pt}
    \begin{tabular}{lccc}
    \toprule
     & Model 1 & Model 2 & Model 3 \\
    \multicolumn{4}{c}{Dependent Variable: Assortativity} \\
    \midrule
    Intercept & 0.392*** & 0.413*** & 0.432*** \\
      & (0.013) & (0.012) & (0.016) \\
    First Night (2-3 February) & -0.053*** & -0.031*** & -0.024 \\
      & (0.011) & (0.009) & (0.014) \\
    First Night:Treatment & -0.143*** & -0.188*** & -0.188*** \\
      & (0.014) & (0.021) & (0.019) \\
    Threshold & 0.027 & -0.016 & -0.042 \\
      & (0.021) & (0.018) & (0.024) \\
    Treatment:Threshold & 0.040* & 0.127*** & 0.127*** \\
      & (0.019) & (0.038) & (0.033) \\
    Time Delta & -0.002* & $<$0.001 & 0.004 \\
      & (0.001) & (0.001) & (0.002) \\
    Time Delta:Treatment &   & -0.005** & -0.005*** \\
      &   & (0.002) & (0.002) \\
    Treatment & -0.155*** & -0.198*** & -0.198*** \\
      & (0.012) & (0.021) & (0.017) \\
    Weekday & -0.008*** & -0.008*** & -0.008*** \\
      & (0.002) & (0.002) & (0.002) \\
    Time Delta ** 2 &   &   & -$<$0.001 \\
      &   &   & ($<$0.001) \\
    Time Delta ** 3 &   &   & -$<$0.001 \\
      &   &   & ($<$0.001) \\
    \midrule
    R-squared & 0.844 & 0.863 & 0.873 \\
    Adjusted R-squared & 0.824 & 0.843 & 0.849 \\
    No. observations & 64 & 64 & 64 \\
    \bottomrule
    \end{tabular}
    \begin{tablenotes}
    \item \textit{Notes:} Standard Errors are in parentheses. Standard Errors are heteroscedasticity and autocorrelation robust (HAC) using 1 lag. \newline
    *** p$<$0.001, ** p$<$0.01, * p$<$0.05.
    \end{tablenotes}
    \label{tab:regression_results_assort_moved_dd}
\end{threeparttable}
\end{center}
\end{table}

\begin {table}[H]
\begin{center}
\begin{threeparttable}
    \caption{RDiT Regression Results for the Assortativity (All)}
    \setlength\extrarowheight{-2pt}
    \begin{tabular}{lccc}
    \toprule
     & Model 1 & Model 2 & Model 3 \\
    \multicolumn{4}{c}{Dependent Variable: Assortativity} \\
    \midrule
    Intercept & 0.951*** & 0.898*** & 0.940*** \\
      & (0.014) & (0.026) & (0.015) \\
    First Night (2-3 February) & -0.273*** & -0.217*** & -0.187*** \\
      & (0.055) & (0.060) & (0.056) \\
    Threshold & -0.165*** & -0.172*** & -0.240*** \\
      & (0.046) & (0.041) & (0.049) \\
    Time Delta & -0.001 & -0.019* & -0.008 \\
      & (0.001) & (0.008) & (0.004) \\
    Time Delta:Threshold & 0.010* & 0.047* & 0.042** \\
      & (0.005) & (0.019) & (0.013) \\
    Weekday & -0.014** & -0.013*** & -0.014*** \\
      & (0.004) & (0.004) & (0.003) \\
    Time Delta ** 2 &   & -0.001* & -0.001** \\
      &   & ($<$0.001) & ($<$0.001) \\
    Time Delta ** 3 &   &   & -$<$0.001** \\
      &   &   & ($<$0.001) \\
    \midrule
    R-squared & 0.864 & 0.901 & 0.925 \\
    Adjusted R-squared & 0.838 & 0.877 & 0.903 \\
    No. observations & 32 & 32 & 32 \\
    \bottomrule
    \end{tabular}
    \begin{tablenotes}
    \item \textit{Notes:} Standard Errors are in parentheses. Standard Errors are heteroscedasticity and autocorrelation robust (HAC) using 1 lag. \newline
    *** p$<$0.001, ** p$<$0.01, * p$<$0.05.
    \end{tablenotes}
    \label{tab:regression_results_assort_all_rdd}
\end{threeparttable}
\end{center}
\end{table}

\begin {table}[H]
\begin{center}
\begin{threeparttable}
    \caption{DiD Regression Results for the Assortativity (All)}
    \setlength\extrarowheight{-2pt}
    \begin{tabular}{lccc}
    \toprule
     & Model 1 & Model 2 & Model 3 \\
    \multicolumn{4}{c}{Dependent Variable: Assortativity} \\
    \midrule
    Intercept & 1.003*** & 0.984*** & 0.997*** \\
      & (0.020) & (0.009) & (0.017) \\
    First Night (2-3 February) & -0.003 & -0.023*** & 0.010 \\
      & (0.018) & (0.006) & (0.024) \\
    First Night:Treatment & -0.340*** & -0.298*** & -0.298*** \\
      & (0.018) & (0.039) & (0.034) \\
    Threshold & -0.049 & -0.008 & -0.050 \\
      & (0.027) & (0.009) & (0.029) \\
    Treatment:Threshold & -0.083*** & -0.164** & -0.164** \\
      & (0.020) & (0.064) & (0.056) \\
    Time Delta & 0.002 & -$<$0.001 & 0.005 \\
      & (0.002) & ($<$0.001) & (0.003) \\
    Time Delta:Treatment &   & 0.005 & 0.005 \\
      &   & (0.003) & (0.003) \\
    Treatment & -0.042*** & -0.003 & -0.003 \\
      & (0.008) & (0.026) & (0.023) \\
    Weekday & -0.009*** & -0.009*** & -0.010*** \\
      & (0.002) & (0.002) & (0.003) \\
    Time Delta ** 2 &   &   & $<$0.001 \\
      &   &   & ($<$0.001) \\
    Time Delta ** 3 &   &   & -$<$0.001 \\
      &   &   & ($<$0.001) \\
    \midrule
    R-squared & 0.850 & 0.865 & 0.887 \\
    Adjusted R-squared & 0.832 & 0.846 & 0.866 \\
    No. observations & 64 & 64 & 64 \\
    \bottomrule
    \end{tabular}
    \begin{tablenotes}
    \item \textit{Notes:} Standard Errors are in parentheses. Standard Errors are heteroscedasticity and autocorrelation robust (HAC) using 1 lag. \newline
    *** p$<$0.001, ** p$<$0.01, * p$<$0.05.
    \end{tablenotes}
    \label{tab:regression_results_assort_all_dd}
\end{threeparttable}
\end{center}
\end{table}

\end{document}